\documentclass[aps,pre,reprint,superscriptaddress,longbibliography]{revtex4-2}

\usepackage{graphicx}
\usepackage{array}
\usepackage{amsmath,amssymb}
\usepackage{xcolor}
\usepackage{hyperref}
\hypersetup{hidelinks, pdftitle={Cohesion from Action-Dependent Perception in Decision-Based Active Matter}, pdfauthor={Julian Giraldo-Barreto and Viktor Holubec}}

\DeclareMathOperator*{\argmax}{arg\,max}

\begin{document}

\title{Cohesion from Action-Dependent Perception in Decision-Based Active Matter}

\author{Julian Giraldo-Barreto}
\email{julian.giraldob@matfyz.cuni.cz}
\author{Viktor Holubec}
\email{viktor.holubec@matfyz.cuni.cz}
\affiliation{Department of Macromolecular Physics, Faculty of Mathematics and Physics, Charles University, 18000 Prague, Czech Republic}

\begin{abstract}
Animals and robots can be modeled as choosing, among a few candidate actions, the one that maximizes a utility built from what they perceive. If the utility sums pairwise scores that involve no positions, such as alignment, the positions of the neighbors can favor an action only if the perceived neighbors depend on the action. For a vision cone of finite range, the first angular harmonic of the cone turns this dependence into a score-weighted pointing interaction toward the neighbors. We simulate agents that align with their $k$ nearest neighbors, at any distance, in a cone along each candidate heading. Since a candidate heading whose cone already holds $k$ agents cannot gain further neighbors, the effective attraction acts only at the edge of the group. The agents form self-confined flocks and swarms in open space. With narrow vision cones, sufficiently large alignment errors, and discouraged turns toward headings without visible neighbors, the groups also form swirls that spontaneously reverse their direction of rotation. With a cone fixed along the heading, as in Vicsek-type models, agents are not confined at any nonzero noise.
\end{abstract}

\maketitle

\section{Introduction}
\label{sec:intro}

Active matter physics~\cite{Vicsek201271,Romanczuk2012,BechingerReview2016,AMRoadmap2025} studies collective organization in systems of self-propelled agents, from bacteria and synthetic microswimmers to bird flocks, human crowds, and intelligent matter~\cite{Ballerini2008,Berdahl2013,Silverberg2013,Kaspar2021}. Most models describe agents through relaxational rules, in which each agent continuously adjusts its state toward a target such as the local average orientation~\cite{Reynolds1987,Vicsek1995,Couzin2005,Shaebani2020,Barberis2016}. Many agents, however, act through discrete events: bacteria switch between runs and tumbles~\cite{BERG1972,Othmer1988}, and animals, pedestrians, and robots choose among a few options~\cite{Ecology,MehdiCrowds2011,AMRoadmap2025,VitelliDecisionMaking,Puri2026}. Decision-based models describe such choices as the maximization of a utility over a finite set of candidate actions~\cite{Schweitzer2003,Charlesworth2019,Turner2023,BH2025}, and thereby represent explicitly what an agent perceives, which actions are open to it, and what it seeks to achieve. Recent examples include particles that condition their motion on local measurements of their environment~\cite{VitelliDecisionMaking} and robotic swarms whose binary decisions integrate a memory of past interactions~\cite{Li2026Memory}.

We ask how the form of a decision rule determines the effective interactions between agents in open space, where alignment alone, as in the Vicsek model, does not keep a finite group together~\cite{Gineli2010,Ginelli2016}, so that models add confinement~\cite{NAGY2007,Albaladejo2023} or an explicit attraction~\cite{Reynolds1987,COUZIN2002,Couzin2005,Caprini2023}. We consider rules whose utility is a sum of pairwise scores over the perceived neighbors. If the scores involve no positions, as for alignment, the positions of the neighbors can favor an action only if the action changes which neighbors are perceived; if it does not, an alignment rule is a Vicsek-type model, and if it does, the decision favors actions that bring well-scoring neighbors into view. For a vision cone of finite range, the first angular harmonic of this preference is the familiar interaction that rewards an agent for pointing toward its neighbors~\cite{COUZIN2002,Gregoire2004,Barberis2016}, here weighted by their scores.

We previously studied a rule of this class with metric neighborhoods~\cite{BH2025}. It produced cohesive flocks without explicit attraction, but since the number of neighbors within a metric range, and with it the reward for recruiting them, grows with the density without bound, the flocks contracted until each agent interacted with most of the others. Here we study scanning alignment: agents with a limited vision cone~\cite{FernandezJuricic2004,Martin2007,LavergneBechingerScience2019,Negi2024} evaluate each candidate heading by their alignment with the $k$ nearest neighbors in the cone along that heading~\cite{Ballerini2008,Gineli2010}. A cone that holds $k$ agents cannot recruit more, so the effective attraction acts only on agents that have a candidate cone with fewer than $k$ agents, at the edge of the group, and the groups do not collapse into dense states. They form self-confined flocks and swarms in open space and, for narrow cones and large turns per decision, swirls whose sense of rotation reverses spontaneously in small groups.

Section~\ref{sec:gen_model} shows how the positions of the neighbors enter additive decision rules. Section~\ref{sec:ea_model} defines the scanning-alignment model and three variants of its decision rule, and derives where and in which form the effective attraction acts. Section~\ref{sec:sim_prot} describes the simulations and observables, Sec.~\ref{sec:results} presents the numerical results, and Sec.~\ref{sec:discussion} concludes. Appendices~\ref{sec:movies}--\ref{sec:refine} provide supplementary information on the generality of the numerical results and describe the supplementary videos.

\section{Decision rules with action-dependent neighborhoods}
\label{sec:gen_model}

Consider $N$ agents, such as self-propelled particles, that act in discrete time. At each step, agent $i$ chooses an action $\alpha$, e.g., a reorientation, from a set $\Omega_i^t$ of admissible actions. The action fixes its next velocity and, possibly, an internal state such as an opinion. The chosen action maximizes a utility,
\begin{equation}
\alpha_i^t=\underset{\alpha\in\Omega_i^t}{\arg\max}\;U_i^t(\alpha),
\label{eq:argmax}
\end{equation}
that depends on the agents with which $i$ would interact if it took action $\alpha$. We consider utilities that are sums of pairwise scores,
\begin{equation}
U_i^t(\alpha)=\sum_{j\neq i} w_{ij}^t(\alpha)\,s_{ij}^t(\alpha),
\label{eq:general_utility}
\end{equation}
where $w_{ij}^t(\alpha)\in\{0,1\}$ indicates whether $j$ belongs to the neighborhood that $i$ perceives, and interacts with, if it takes action $\alpha$, and $s_{ij}^t(\alpha)$ is the score of that interaction, e.g., the degree of alignment or agreement. Scores need be neither symmetric nor nonnegative.

The decision depends only on differences of the utility between candidate actions, which split into two parts,
\begin{align}
U_i^t(\alpha)-U_i^t(\alpha')={}&\sum_{j} w_{ij}^t(\alpha')\big[s_{ij}^t(\alpha)-s_{ij}^t(\alpha')\big]\nonumber\\
&+\sum_{j}\big[w_{ij}^t(\alpha)-w_{ij}^t(\alpha')\big]\,s_{ij}^t(\alpha).
\label{eq:split}
\end{align}
The first part, which we call the rescoring term, compares the scores of the two actions against the same neighbors. For an alignment score, it captures ordinary alignment. The second, the recruitment term, accounts for changes in the neighbor sets. If the neighborhood is independent of the action, so that $w_{ij}^t(\alpha)=w_{ij}^t$, the recruitment term vanishes. If the neighborhood does depend on the action, the recruitment term favors actions that recruit neighbors with positive scores and disfavors actions that recruit neighbors with negative ones. For predominantly positive scores, it thus favors actions that bring more neighbors into view, which is a cohesive bias when the perceived neighborhood lies along the direction of motion, as for a vision cone. 

Positions can also affect the decision through the admissible set $\Omega_i^t$. For example, this set may contain only actions that recruit at least one neighbor.

\section{Scanning-alignment model}
\label{sec:ea_model}

We now consider agents as point particles without excluded volume, moving in two-dimensional open space at constant speed $v_0$. The position $\mathbf r_i^t$ and orientation $\theta_i^t$ of agent $i$ evolve in discrete time according to
\begin{align}
\mathbf r_i^{t+\Delta t}
&=
\mathbf r_i^t
+
\mathbf v_i^{t+\Delta t}\,\Delta t,
\label{eq:Pos_upd}\\
\theta_i^{t+\Delta t}
&=
\theta_i^t+\alpha_i^t+\xi_i^t,
\label{eq:Ang_upd}
\end{align}
where $\Delta t$ is the time step,
\begin{equation}
\mathbf v_i^{t+\Delta t}
=
v_0
\bigl(
\cos\theta_i^{t+\Delta t},
\sin\theta_i^{t+\Delta t}
\bigr)
\label{eq:speed_ALModel}
\end{equation}
is the velocity after the update, and the angular noise $\xi_i^t$ is drawn independently from the uniform distribution on $[-\eta\pi,\eta\pi]$, with noise amplitude $\eta\in[0,1]$. The candidate actions are reorientations applied before the noise. The simplest choice for a motile agent is to turn left, turn right, or keep its heading,
\begin{equation}
\alpha\in
\{-\alpha_d,\,0,\,\alpha_d\}.
\label{eq:alpha}
\end{equation}
We call these reorientations candidate actions, or simply candidates, the corresponding headings $\theta_i^t+\alpha$ candidate headings, and the vision cones placed along them candidate cones; $\Omega_i^t$ contains the candidates that are admissible.
The score measures the alignment of the intended velocity
\begin{equation}
\tilde{\mathbf v}_i^t(\alpha)
=
v_0
\bigl(
\cos(\theta_i^t+\alpha),
\sin(\theta_i^t+\alpha)
\bigr),
\end{equation}
with the current velocity of neighbor $j$,
\begin{equation}
s_{ij}^t(\alpha)
=
a+
\tilde{\mathbf v}_i^t(\alpha)\cdot\mathbf v_j^t,
\label{eq:score_general}
\end{equation}
where the constant $a$ is specified in Sec.~\ref{sec:EAM}. The agent chooses its action according to Eq.~\eqref{eq:argmax}, with the utility~\eqref{eq:general_utility}. To preserve chiral symmetry, a tie between the left and right turns alone is broken at random with equal probability. Other ties are resolved in favor of the straight direction. 

We denote the number of recruited neighbors by $N_i^t(\alpha)=\sum_j w_{ij}^t(\alpha)$ and, for $N_i^t(\alpha)>0$, we define the mean velocity of the recruited neighbors, $\mathbf V_i^t(\alpha)=N_i^t(\alpha)^{-1}\sum_j w_{ij}^t(\alpha)\,\mathbf v_j^t$. The utility of an action with $N_i^t(\alpha)>0$ is
\begin{equation}
U_i^t(\alpha)
=
N_i^t(\alpha)\,\left(a + 
\tilde{\mathbf v}^t_i(\alpha)\cdot
\mathbf V_i^t(\alpha)\right).
\label{eq:factorized}
\end{equation}
If the neighborhood does not depend on the action, $w^t_{ij}(\alpha)=w^t_{ij}$, the factor $N_i^t$ is common to all candidates $\alpha$, and the rule selects the candidate velocity best aligned with the mean velocity $\mathbf V_i^t$ of the neighbors,
\begin{equation}
\alpha_i^t
=
\underset{\alpha\in\Omega_i^t}{\arg\max}\,
\tilde{\mathbf v}^t_i(\alpha)\cdot\mathbf V_i^t,
\label{eq:vicsek_limit}
\end{equation}
a Vicsek-type alignment rule~\cite{Vicsek1995} with a discrete set of reorientations, for any choice of neighborhood.

\subsection{Rigid and scanning agents}
\label{sec:SEAM}

\begin{figure}
\centering
\includegraphics[width=1.0\columnwidth]{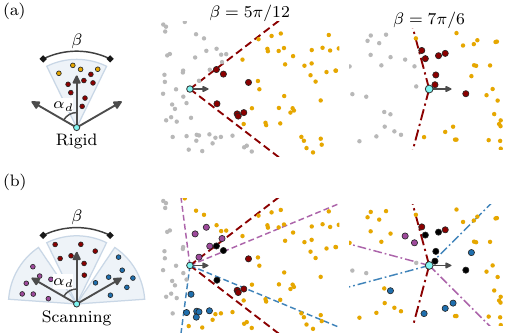}
\caption{\textbf{Rigid and scanning agents.}
(a) \emph{Rigid agents}: the interaction neighborhood is determined by the current heading and consists of the $k$ nearest agents within a vision cone of opening angle $\beta$ centered on the heading. The selected reorientation $\alpha\in\{-\alpha_d,0,\alpha_d\}$ maximizes alignment with this fixed neighbor set, a variant of the Vicsek model. (b) \emph{Scanning agents}: a separate cone is placed along each candidate heading, and the selected reorientation maximizes the summed alignment score $U_i^t(\alpha)$ with the neighbor set of its cone. Left: schematic with cones of finite range for clarity. Right: neighbor sets of a focal agent (cyan; the arrow gives its heading) in a random configuration, for $k=7$, $\alpha_d=\pi/3$, and two cone angles. Lines mark the boundaries of the cones along the current heading (dark red) and, in (b), along the left (purple) and right (blue) candidate headings. Gray dots denote agents outside every cone, yellow dots agents inside a cone but not among its $k$ nearest, dark red, purple, and blue dots the $k$ nearest agents in the corresponding cone, and black dots agents among the $k$ nearest of more than one cone.}
\label{fig:fig1}
\end{figure}

The neighborhoods used here are topological~\cite{Ballerini2008,Gineli2010} and restricted to a vision cone of opening angle $\beta<2\pi$~\cite{FernandezJuricic2004,Martin2007,LavergneBechingerScience2019,Negi2024}. For each candidate $\alpha$, $w^t_{ij}(\alpha)$ selects the $k$ nearest agents within the cone, or all agents in the cone if it contains fewer than $k$. The cones have no metric cutoff, as in metric-free models~\cite{Ballerini2008,Gineli2010}. Appendix~\ref{sec:range} shows what changes with a finite perception range. We call agents \emph{rigid} [Fig.~\ref{fig:fig1}(a)] when the cone is fixed by the current heading, so that the neighborhood is the same for all candidates and the dynamics reduces to the Vicsek-type rule~\eqref{eq:vicsek_limit}. Rigid agents are closely related to vision-restricted Vicsek models with metric neighborhoods in periodic domains, in which a limited field of view shifts the flocking transition~\cite{Nguyen2015,Durve2016} and, with delayed response, produces condensation without attraction~\cite{Durve2018}. We call agents \emph{scanning} [Fig.~\ref{fig:fig1}(b)] when they evaluate a separate cone along each candidate heading, so that their neighborhood depends on the action whenever $\beta<2\pi$.

\subsection{Three variants of the decision rule}
\label{sec:EAM}

The three variants studied (Table~\ref{tab:variants}) share the score~\eqref{eq:score_general} and differ in the offset $a$ and in whether empty candidate cones compete: (i)~the \emph{baseline} model, with $a=0$ and all candidate cones admitted; (ii)~the \emph{nonnegative score} model, with $a=v_0^2$, for which $s_{ij}^t(\alpha)\geq0$ because $|\tilde{\mathbf v}_i^t(\alpha)|=|\mathbf v_j^t|=v_0$; and (iii)~the \emph{visible-neighbor} model, with $a=0$ and $\Omega_i^t=\{\alpha\in\{-\alpha_d,0,\alpha_d\}:N_i^t(\alpha)>0\}$, in which an agent that perceives no neighbor along any candidate cone keeps its heading up to the noise. If every candidate cone of an agent holds at least $k$ agents, each recruits exactly $k$, so that the offset adds the same constant $ak$ to all candidates $\alpha$ and no candidate cone is empty. The three variants then make the same decision, as they do for rigid agents. 

They differ only for agents with an under-filled candidate cone with fewer than $k$ agents, in two respects. First, with $a=0$, a recruited neighbor that moves against the candidate heading lowers the utility, whereas with $a=v_0^2$ every recruited neighbor raises it. Second, an empty candidate cone, with $U_i^t=0$, is chosen in variant~(i) whenever every recruiting candidate cone has $U_i^t<0$, i.e., the agent's neighbors tend to move in the opposite direction. In variant~(ii), the empty cone can at most tie, since every recruiting candidate has $U_i^t\geq0$. And in variant~(iii), the empty cone is excluded, which is the second way in which the neighborhood enters the decision (Sec.~\ref{sec:gen_model}). All three variants produce confined states. The main text presents the visible-neighbor model~(iii), which we consider the most natural choice for a social task, since a heading along which no conspecific is visible offers no basis for coordination; variants~(i) and~(ii) are documented in Appendix~\ref{ssec:score_variants}.

\begin{table}
\caption{The three decision rules: (i) baseline, (ii) nonnegative score, and (iii) visible-neighbor model. All three make the same decision wherever every candidate cone holds at least $k$ agents, and for rigid agents. ``Empty candidates'' states how candidates $\alpha$ with $N_i^t(\alpha)=0$ enter the optimization, ``Sign of scores'' whether a recruited neighbor can lower the utility, and the last column whether swirling states occur (Appendix~\ref{ssec:score_variants}).}
\label{tab:variants}
\begin{ruledtabular}
\begin{tabular}{lcccc}
Model & $a$ & Empty candidates & Sign of scores & Swirls\\
\hline
(i) & $0$ & can be chosen & signed & no\\
(ii) & $v_0^2$ & can at most tie & nonnegative & yes\\
(iii) & $0$ & excluded & signed & yes\\
\end{tabular}
\end{ruledtabular}
\end{table}

\subsection{Effective attraction}
\label{ssec:pointing}

In this subsection, we drop the time index and set $v_0=1$, so that $a$ is measured in units of $v_0^2$ and the candidate heading $\tilde{\mathbf v}_i(\alpha)$ is a unit vector. 

A candidate action can be rewarded for the number of neighbors it recruits only if some candidate cone of the agent holds fewer than $k$ agents. Because the cones have no metric cutoff, each candidate cone is an unbounded wedge, which holds fewer than $k$ agents only near the edge of a group [Fig.~\ref{fig:edge} in Appendix~\ref{ssec:boundary}]. Consider an agent on the boundary of a convex group. For a narrow cone, $\beta<\pi$, the wedge for a candidate turn normal to the group contains no agents for any group size. On the other hand, for a broad cone, $\beta>\pi$, the wedge reaches $(\beta-\pi)/2$ behind the group tangent on either side. For an agent on the edge of a uniform disc of radius $R_{\rm d}$ holding $N$ agents, the wedge along the outward normal then covers two circular segments with central angle $\beta-\pi$, of total area $R_{\rm d}^2(\beta-\pi+\sin\beta)$, which hold $(N/\pi)(\beta-\pi+\sin\beta)$ agents. We thus obtain the condition 
\begin{equation} 
(N/\pi)(\beta-\pi+\sin\beta) < k, 
\label{eq:broad_rim_k} 
\end{equation} 
which must be satisfied for such circular flocks to remain cohesive. Thus, in compact broad-cone groups with $N > k\pi(\beta-\pi+\sin\beta)^{-1}$, under-filled cones and hence cohesion can occur only for agents that stray beyond the group's edge or are located at sparse or irregular edges. Appendix~\ref{ssec:boundary} tests this localization, and Appendix~\ref{sec:kbeta} demonstrates the decrease in cohesive bias with increasing $\beta$ and decreasing $k$ for fixed $N$, as predicted by Eq.~\eqref{eq:broad_rim_k}.

For a cone of finite range $R$ that recruits all agents within it (a metric cone not simulated here), the effective attraction can be written as an explicit interaction, which allows a comparison with models that add attraction as a separate term. Neighbor $j$ then lies in the cone of candidate $\alpha$ if $r_{ij} = |\mathbf{r}_i - \mathbf{r}_j| <R$ and if its bearing $\psi_{ij}(\alpha)$ relative to the candidate heading, given by $\cos\psi_{ij}(\alpha)=\tilde{\mathbf v}_i(\alpha)\cdot\hat{\mathbf r}_{ij}$ with $\hat{\mathbf r}_{ij} = (\mathbf{r}_j - \mathbf{r}_i)/r_{ij}$ the unit vector from $i$ to $j$, satisfies $|\psi_{ij}(\alpha)|\leq\beta/2$. The cone indicator is $w_{ij}(\alpha)=\Theta(R-r_{ij})\,\Theta\big(\beta/2-|\psi_{ij}(\alpha)|\big)$, with $\Theta$ the Heaviside step function. Its angular part is an even, $2\pi$-periodic function of the bearing, and expanding it in a cosine Fourier series yields
\begin{align}
w_{ij}(\alpha)={}&\Theta(R-r_{ij})\Big[\frac{\beta}{2\pi}+\frac{2}{\pi}\sin\frac{\beta}{2}\;\tilde{\mathbf v}_i(\alpha)\cdot\hat{\mathbf r}_{ij}\nonumber\\
&+\sum_{n\geq2}\frac{2}{n\pi}\sin\frac{n\beta}{2}\,\cos n\psi_{ij}(\alpha)\Big].
\label{eq:harmonics}
\end{align}
Inserting Eq.~\eqref{eq:harmonics} and the score~\eqref{eq:score_general} into the utility~\eqref{eq:general_utility} and dropping the term $(\beta/2\pi)\,a\sum_j\Theta(R-r_{ij})$, which is the same for all candidates $\alpha$, gives
\begin{equation}
U_i(\alpha)=\frac{\beta}{2\pi} \tilde{\mathbf v}_i(\alpha)\cdot\Big[\sum_j\mathbf v_j
+\lambda_\beta\sum_j s_{ij}(\alpha)\,\hat{\mathbf r}_{ij}\Big]+\dots,
\label{eq:pointing}
\end{equation}
with the coupling $\lambda_\beta=4\sin(\beta/2)/\beta$. The sums run over all agents within $R$ and the dots stand for the harmonics $n\geq2$. The first term is Vicsek alignment with all agents within $R$. The second is a pointing interaction: it rewards the candidate heading for pointing toward each neighbor, in proportion to the score of that neighbor. We note that the higher harmonics are generally not small, since the coefficient of the second harmonic is $\cos(\beta/2)$ times that of the first. Consequently, truncating Eq.~\eqref{eq:pointing} after the first harmonic is not a controlled approximation.

For scores that do not depend on the candidate heading, the utility~\eqref{eq:pointing} is maximized by the heading parallel to the expression in the bracket, as in explicit models~\cite{Gregoire2004,Barberis2016}. Indeed, the deterministic part of the update of Gr\'egoire and Chat\'e~\cite{Gregoire2004} selects among all headings the one that maximizes $\hat{\mathbf e}\cdot(\sum_j\mathbf v_j+\lambda\sum_jf_{ij}\hat{\mathbf r}_{ij})$, with a distance-dependent attraction--repulsion $f_{ij}$ in place of $s_{ij}$ and a strength $\lambda$. Here, only the offset $a$ of the score is of this kind; its alignment part $\tilde{\mathbf v}_i(\alpha)\cdot\mathbf v_j$ strengthens the pointing toward neighbors that move along the candidate heading and weakens it toward those that move against it. 

For the topological cones of the present model, the same derivation holds with $R$ replaced by $R_i(\alpha)$, the distance to the $k$th nearest agent in the cone of $\alpha$ ($R_i(\alpha)=\infty$ if the cone holds fewer than $k$ agents). The sums in Eq.~\eqref{eq:pointing} then depend on $\alpha$, and the dropped term $(\beta/2\pi)\,a\sum_j\Theta\big(R_i(\alpha)-r_{ij}\big)$ must in general be retained. For $a=0$, it vanishes, and the first harmonic still gives a score-weighted pointing interaction. For an under-filled candidate cone, the sums in Eq.~\eqref{eq:pointing} run over all agents. A heading away from the group, whose cone is empty, then has $\tilde{\mathbf v}_i(\alpha)\cdot\hat{\mathbf r}_{ij}<0$ for nearly all $j$ and is disfavored as long as these agents move along it, $s_{ij}(\alpha)>0$, whereas headings toward the group are favored. Such candidate cones occur for agents at or beyond the rim of the group.

\section{Simulations and observables}
\label{sec:sim_prot}

Time and length are measured in units of $\Delta t$ and $v_0\Delta t$, in which $v_0$ drops out of the dynamics; the control parameters are the noise amplitude $\eta$, the reorientation angle $\alpha_d$, the cone angle $\beta$, and the neighbor number $k$. Except for a check with random initial orientations (Appendix~\ref{sec:zeronoise}) and the runs of Appendices~\ref{sec:range} and~\ref{sec:refine}, which continue from groups formed in this way, all runs start from a fully polarized state, with a common random heading, at random positions in a square of side $L=100\sqrt{N/200}$, so that the initial density does not depend on $N$. All agents are updated synchronously, in unbounded two-dimensional space. Unless stated otherwise, $N=200$, each parameter point is simulated in 25 independent replicas of $t_f=10^5$ steps, and statistics are taken over $t\in[t_f/3,t_f]$.

\paragraph*{Parameters.}
We use $k=7$, which lies at the onset of the range in which confinement no longer depends on $k$ (Appendix~\ref{sec:kbeta}), and two cone angles representative of narrow ($\beta=5\pi/12$) and broad ($\beta=7\pi/6$) fields of view, which leaves the $(\eta,\alpha_d)$ plane to be explored. The dependence of the dynamics on $\beta$ and $k$ is studied in Appendix~\ref{sec:kbeta}. For the narrow cone, a single neighbor, $k=1$, already confines the groups in nearly the same region, but without swirls; for the broad cone, it does not, in accord with Eq.~\eqref{eq:broad_rim_k}. Results for a larger number of possible reorientations than three are presented in Appendix~\ref{sec:refine}.

\paragraph*{Observables.}
We use positions and velocities in the center-of-mass frame,
\begin{equation}
\bar{\mathbf r}_i^t
= \mathbf r_i^t - \mathbf R_{\rm c.m.}^t,
\qquad
\mathbf R_{\rm c.m.}^t
= \frac{1}{N}\sum_{i=1}^{N}\mathbf r_i^t,
\label{eq:CM_position}
\end{equation}
and
\begin{equation}
\bar{\mathbf v}_i^t
= \mathbf v_i^t - \mathbf V_{\rm c.m.}^t,
\qquad
\mathbf V_{\rm c.m.}^t
= \frac{1}{N}\sum_{i=1}^{N}\mathbf v_i^t.
\label{eq:CM_velocity}
\end{equation}
The group extent,
\begin{equation}
\delta_{\rm c.m.}^t
= \sqrt{\frac{1}{N}
\sum_{i=1}^{N}
\left|
\bar{\mathbf r}_i^t
\right|^2},
\label{eq:SysSize}
\end{equation}
weights distant agents strongly, a single agent at distance $D\gg\delta^t_{\rm c.m.}$ raising it to about $D/\sqrt N$, and therefore registers escaping agents promptly. For a uniform disc of radius $R_{\rm d}$, $\delta^t_{\rm c.m.}=R_{\rm d}/\sqrt2$. Collective rotation is measured by the angular velocity about the center of mass,
\begin{equation}
\omega^t
= \frac{1}{N}
\sum_{i=1}^{N}
\frac{
\bar{\mathbf r}_i^t\times \bar{\mathbf v}_i^t
}{
|\bar{\mathbf r}_i^t|^2
},
\label{eq:AngularVelocity}
\end{equation}
where in two dimensions the cross product is a scalar, and orientational order by the polarization
\begin{equation}
\phi^t
= \frac{1}{v_0}
\left|
\mathbf V_{\rm c.m.}^t
\right|.
\label{eq:Polarization}
\end{equation}
Flocks have $\phi^t\approx1$ and $\omega^t\approx0$, swarms $\phi^t\approx0$ and $\omega^t\approx0$, and swirls $\phi^t\approx0$ and $|\omega^t|>0$. Since both senses of rotation are equally likely, $\omega^t$ averages to zero over the replicas, and we use $|\omega^t|$. 

Since $\omega^t$ weights the tangential velocity of each agent by the inverse of its distance from the center of mass, an agent that passes close to the center of mass produces a spike of $\omega^t$ that lasts one step. Such spikes would be counted as reversals of the sense of rotation. To detect reversals (Sec.~\ref{sec:swirling}), we therefore use the bounded milling parameter
\begin{equation}
M^t=\frac{1}{N}\sum_{i=1}^{N}\frac{\bar{\mathbf r}_i^t\times\mathbf v_i^t}{|\bar{\mathbf r}_i^t|\,v_0},
\label{eq:mill}
\end{equation}
which lies in $[-1,1]$ and equals $\pm1$ for a perfect mill~\cite{Calovi2014}. For a group that translates rigidly with velocity $\mathbf V^t_{\rm c.m.}$, $M^t=(N^{-1}\sum_i\bar{\mathbf r}_i^t/|\bar{\mathbf r}_i^t|)\times\mathbf V^t_{\rm c.m.}/v_0$, which vanishes if the group is symmetric about its direction of motion, but not in general. We therefore use $M^t$ to study the reversals only. 

Angular velocity and extent of a rotating group are related. If the agents move tangentially with a speed $u$ that does not depend on their distance $r$ from the center of mass, then $\omega^t=u\,\overline{r^{-1}}$ and $\delta^t_{\rm c.m.}=(\overline{r^2})^{1/2}$, where the bar denotes the average over the agents, so that
\begin{equation}
\omega^t\delta^t_{\rm c.m.}=c\,u,\qquad c=\overline{r^{-1}}\,\big(\overline{r^2}\big)^{1/2}\geq1.
\label{eq:omegadelta}
\end{equation}
The factor $c$ depends only on the radial distribution of the agents, and $c\geq1$ because $\overline{r^{-1}}\geq1/\bar r$ and $(\overline{r^2})^{1/2}\geq\bar r$. For a thin ring of radius $r_0$, $\omega^t=u/r_0$, $\delta^t_{\rm c.m.}=r_0$, and $c=1$; for a uniform disc of radius $R_{\rm d}$, $\omega^t=2u/R_{\rm d}$ and $c=\sqrt2$; and for a Gaussian cloud, $c=\sqrt\pi$. For a group without net motion, the mean tangential speed is $u=v_0|M^t|$. Equation~\eqref{eq:omegadelta} serves below as a check on the measured angular velocities and extents of the swirls.

All these observables are time dependent. In the following, averages over replicas and over time are denoted by $\langle X^t\rangle_r$ and $\langle X\rangle_t$, respectively, while the average over both replicas and time is denoted by $\langle X\rangle_{r,t}$.

\section{Results for the visible-neighbor model}
\label{sec:results}

Unless stated otherwise, the results below are for scanning agents in the visible-neighbor model~(iii) with $k=7$ and the protocol of Sec.~\ref{sec:sim_prot}. Averages $\langle\cdot\rangle_{r,t}$ are taken over the statistics window and over the replicas that did not disperse (Sec.~\ref{sec:cohesive}). Figure captions state only departures from these defaults.

\subsection{Stability of confined states}
\label{sec:cohesive}

\begin{figure}[t]
\centering
\includegraphics[width=1.0\columnwidth]{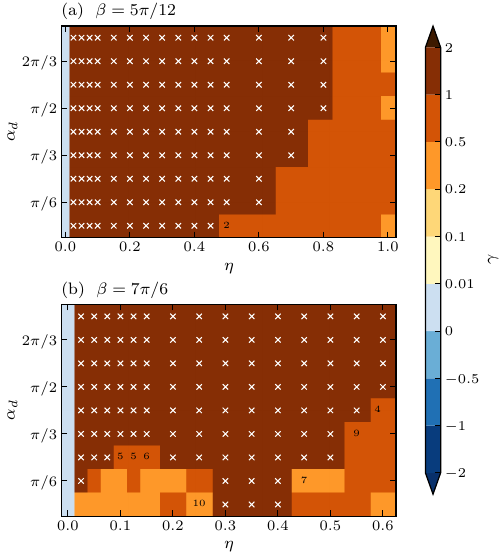}
\caption{\textbf{Stability diagram for rigid agents}, for which the three variants of Table~\ref{tab:variants} make identical decisions.
Exponent $\gamma$ of the late-time growth of the group extent, $\delta^t_{\rm c.m.}\sim t^\gamma$, fitted over $t\in[t_f/10,t_f]$ to the lower quartiles of the extent in ten intervals (Sec.~\ref{sec:cohesive}), as a function of the noise amplitude $\eta$ and the reorientation angle $\alpha_d$ for the vision-cone angles $\beta=5\pi/12$ (a) and $\beta=7\pi/6$ (b). Points at which at least half of the replicas dispersed are shown in the band $1<\gamma<2$ and marked by crosses; elsewhere, numbers give how many replicas dispersed after $t_f/10$. At vanishing noise, the polarized initial state persists unchanged.}
\label{fig:Sfig5}
\end{figure}

\begin{figure*}[t]
\centering
\includegraphics[width=1.0\textwidth]{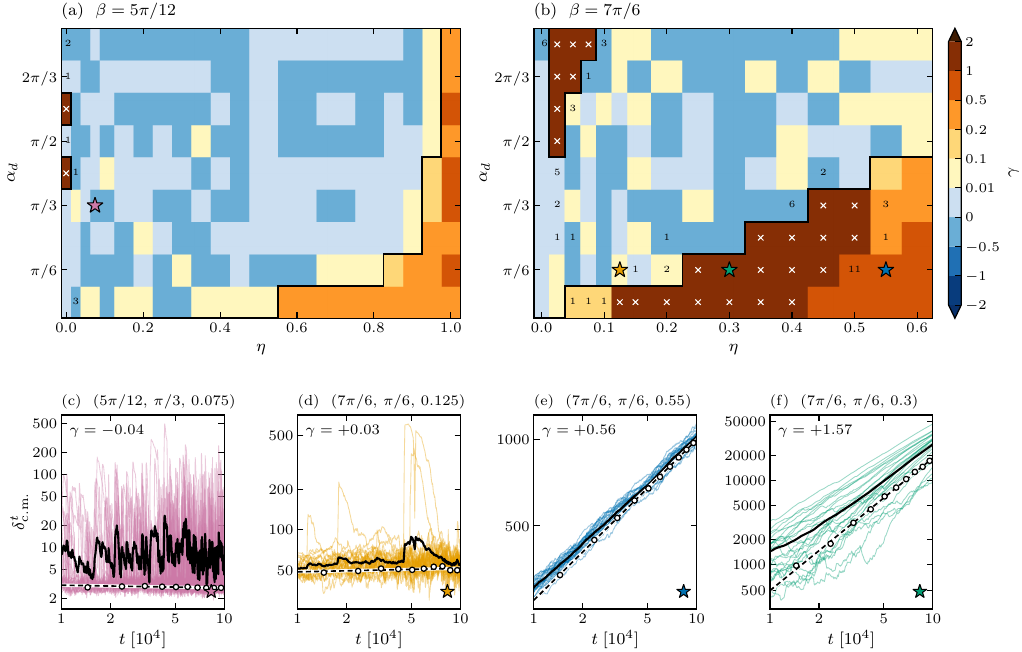}
\caption{\textbf{Stability diagram and representative trajectories of the group extent for scanning agents.}
(a,b) Exponent $\gamma$, crosses, and numbers as in Fig.~\ref{fig:Sfig5}, for $\beta=5\pi/12$ (a) and $\beta=7\pi/6$ (b); black lines enclose the points classified as confined. (c)--(f) Group extent of 25 replicas (thin lines) and their average (thick line) over $t\in[t_f/10,t_f]$, at the parameters $(\beta,\alpha_d,\eta)$ marked by stars in (a,b); circles show the lower quartile of the extent in ten intervals, and the dashed line the robust fit $A\,t^\gamma$ to them. Panels (c)--(e) show the runs of the diagram, so that $\gamma$ is the value colored in (a,b); panel (f) shows separate runs without stopping, since all runs of the diagram at this point are stopped.
(c) A narrow-cone group that repeatedly splits and merges again; the excursions of single replicas dominate the average but not the lower quartile, and the point is confined. (d) A broad-cone group whose lower quartile grows slowly, with $\gamma\approx0.03$ below the threshold, while single replicas make large excursions. (e) Diffusive spreading, $\gamma\approx1/2$. (f) Fast spreading, $\gamma\approx1.6$.}
\label{fig:fig4}
\end{figure*}

We call a state confined, or self-confined, since no walls or external potentials act, if its extent stops growing at long times, up to fluctuations. This is what we mean by cohesion. Because diffusive spreading can be slow, a small extent over a finite observation window does not imply confinement, and we classify states by the growth of $\delta_{\rm c.m.}^t$ rather than by its value, through the exponent of the late-time scaling
\begin{equation}
\delta_{\rm c.m.}^t \sim t^\gamma.
\label{eq:gamma}
\end{equation}
Confined states have $\gamma\lesssim0$, diffusive spreading gives $\gamma=1/2$ [Fig.~\ref{fig:fig4}(e)], and ballistic spreading $\gamma=1$; larger values, as in Fig.~\ref{fig:fig4}(f), reflect replicas that break up at different times. In finite runs, $\gamma$ is a finite-time diagnostic: confined states that relax slowly give small positive values [Fig.~\ref{fig:fig4}(d)], and groups that split temporarily and merge again give excursions of either sign, which dominate the replica average [Fig.~\ref{fig:fig4}(c)].

We therefore fit $\gamma$ in a way that is insensitive to temporary splitting. We divide the last decade of the run, $t\in[t_f/10,t_f]$, into ten successive intervals of equal length and take in each the lower quartile of $\delta^t_{\rm c.m.}$ over the replicas and times. Since temporary splits are present but carry little weight in stable configurations, this quantile is set by the unsplit configurations and so measures the extent of the group when it is not split. The exponent $\gamma$ is the robust (Theil--Sen) slope~\cite{Theil1950,Sen1968} of $\ln \delta_{\rm c.m.}^t$ versus $\ln t$. A replica counts as dispersed, and is stopped, once $\delta^t_{\rm c.m.}/\sqrt N>100$, about 35 times the initial value. Below, a parameter point is deemed confined if $\gamma \le 0.1$ and fewer than half of the corresponding replicas disperse.

Rigid agents are not confined at any nonzero noise amplitude for arbitrary reorientation angle (Fig.~\ref{fig:Sfig5}) and vision cone angle $\beta$ (Appendix~\ref{sec:kbeta}), consistent with the absence of cohesion in topological Vicsek-type alignment~\cite{Gineli2010}. The unbounded interaction range hence does not by itself produce confinement.

Scanning agents are confined in extended regions for both cone angles (Fig.~\ref{fig:fig4}). The confined region ends at a noise amplitude that increases with $\alpha_d$, from $\eta=0.5$ to $\eta\approx0.9$ for the narrow cone and from $\eta=0.025$ to at least $0.6$ for the broad cone. Dispersal inside the confined region is rare. After the group has formed, $t>t_f/10$, it occurs in at most 6 of the 25 replicas, at 18 of the 260 confined points [numbers in Figs.~\ref{fig:fig4}(a) and~\ref{fig:fig4}(b)]. Earlier dispersals, at 9 confined points, come from the initial condition. At weak noise, a group that disperses typically splits into two or three polarized subflocks that move apart. At the noisy edge of the broad-cone region, it dissolves into many fragments.

Which of the two channels of Sec.~\ref{sec:gen_model} confines the groups? In variant~(iii), both act: the recruitment term and the exclusion of empty candidate cones. In variant~(i), in which empty candidate cones compete with recruiting ones, the recruitment term is the only channel, and the confined regions are nearly the same: on the coarser grid of Appendix~\ref{ssec:score_variants}, they hold 85 of the 108 narrow-cone points and 53 of the 72 broad-cone points with $\eta\leq0.6$, against 92 and 53 for variant~(iii). Conversely, scoring every candidate turn with the rigid neighborhood while still excluding empty candidate cones leaves only the second channel. Over the whole narrow-cone diagram, this control confines the groups at 134 of the 157 points at which the model confines them, and at no other point [Fig.~\ref{fig:c3}(c)]. For the broad cone, whose candidate cones are rarely empty (Fig.~\ref{fig:edge}), it confines the groups at only 22 of the 103 points, nine of them at vanishing noise, where the polarized initial state persists [Fig.~\ref{fig:c3}(d)]. Each channel can thus confine on its own, the second one only for the narrow cone, and both require the neighborhood to depend on the action. 

Consistent with this, replacing the utility score $U_i^t$ by the average score $S_i^t = U_i^t/N_i^t$ eliminates confinement of the broad-cone flock but preserves confinement of the narrow-cone states in variant~(iii) (Appendix~\ref{sec:Scontrol}). Interestingly, in variants~(i) and (ii), which differ only by an additive constant in the candidate scores and therefore yield identical decisions, flocking remains stable for narrow cones even though both confining channels appear to be absent. The reason is that turns away from the group often face empty cones, which score $S_i^t=a$, whereas turns toward the aligned group score higher. Consequently, a weak remnant of the first cohesion channel survives the normalization. In contrast, alignment to the average orientation within the vision cone, as implemented for rigid agents, does not introduce such a bias. Swarming and swirling are unstable in these models.
 
The classification hardly depends on the initial condition (Appendix~\ref{sec:zeronoise}). However, it does depend on the unbounded reach of the cones. With a finite perception range, tested for the swirl, the swarm, and the broad-cone flock of Sec.~\ref{ssec:largerN} at $N=200$, the swarm keeps its agents for ranges down to about its radius $R_{\rm d}$, the swirl only for about $14R_{\rm d}$, and the broad-cone flock splits at all ranges tested, up to about $6R_{\rm d}$ (Appendix~\ref{sec:range}). The dependence on group size, the number of topological neighbors $k$, and the vision-cone angle $\beta$ follows Eq.~\eqref{eq:broad_rim_k} and is discussed in Sec.~\ref{ssec:largerN}.

\begin{figure*}[t]
\centering
\includegraphics[width=1.0\textwidth]{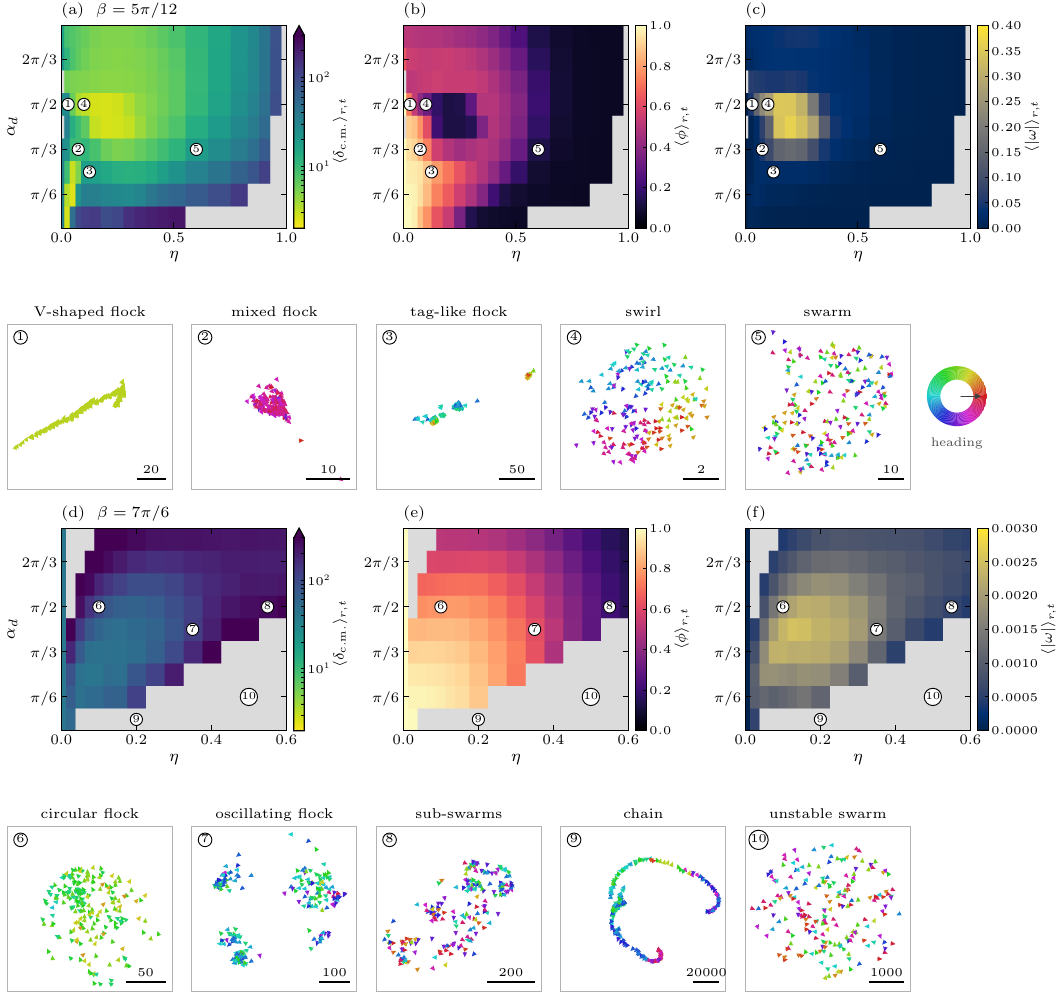}
\caption{\textbf{State diagram of scanning agents.}
(a)--(c) Time- and replica-averaged group extent $\langle\delta_{\rm c.m.}\rangle_{r,t}$, polarization $\langle\phi\rangle_{r,t}$, and absolute angular velocity $\langle|\omega|\rangle_{r,t}$ for the narrow cone, $\beta=5\pi/12$, as functions of the noise amplitude $\eta$ and the reorientation angle $\alpha_d$. (d)--(f) the same for the broad cone, $\beta=7\pi/6$. Points classified as unconfined in Fig.~\ref{fig:fig4} are gray. The noise ranges and the scales of $\langle|\omega|\rangle_{r,t}$ differ between the two cones. Snapshots 1--10 show single configurations at $t=t_f$ from separate runs (no stopping), in the center-of-mass frame, at the parameters marked by the numbered circles. Agents are drawn as arrows colored by their heading (color wheel), and the scale bars give lengths in units of $v_0\Delta t$.}
\label{fig:fig6}
\end{figure*}

\subsection{State diagram}
\label{sec:phase}

Within the confined region, scanning agents form three classes of states: polarized flocks, disordered swarms, and swirls (milling or vortex states). The same classes organize the phase diagrams of animal-group models with explicit attraction and alignment zones~\cite{COUZIN2002,Calovi2014} and are observed in schooling fish~\cite{Tunstrom2013}. Figure~\ref{fig:fig6} shows their organization in the $(\eta,\alpha_d)$ plane, together with snapshots of the states named below.

For the narrow cone ($\beta=5\pi/12$), all three classes occur [Figs.~\ref{fig:fig6}(a)--\ref{fig:fig6}(c)]. At low noise, the group forms polarized flocks. At vanishing noise, they can be V-shaped with unequal arm lengths (snapshot~1; Appendix~\ref{sec:zeronoise}). At higher noise, the flocks alternate between polarized and rotating organization and split and merge again [mixed flocks, snapshot~2 and Fig.~\ref{fig:fig4}(c)], or their polarized clusters repeatedly fragment and reassemble (tag-like flocks, snapshot~3). At large noise, global order is lost while confinement persists, and the group forms a swarm (snapshot~5) whose extent grows with $\eta$ until confinement is lost, at $\eta\approx0.9$ for $\alpha_d\geq\pi/4$. For small $\alpha_d$, the polarization decreases continuously with noise, and the groups are larger.

For $5\pi/12\le \alpha_d \le\pi/2$, i.e., turns of $75^\circ$ to $90^\circ$ per decision, the group forms a compact swirl (snapshot~4) at intermediate noise. For larger turns, the groups at low and intermediate noise stay compact and partly polarized, $\langle\phi\rangle_{r,t}\approx0.5$, but hardly rotate. The swirls are relatively small and dense. At $(\alpha_d,\eta)=(\pi/2,0.2)$, $\langle\delta_{\rm c.m.}\rangle_{r,t}=2.7$, i.e., $R_{\rm d}\approx3.8$ step lengths, and in the swirl at $\eta=0.1$, the median distance between nearest neighbors is about $0.2$ step lengths, so that an agent moves more than four such distances per decision. Their angular velocity and extent agree with Eq.~\eqref{eq:omegadelta}: in the configurations with $|M^t|>0.5$, the ratio $|\omega^t|\delta^t_{\rm c.m.}/(v_0|M^t|)$ has a median of $1.26$ for $0.1\leq\eta\leq0.3$, between the values $1$ for a ring and $\sqrt2$ for a uniform disc, and close to the value $c=1.27$ obtained from the radial positions of the agents, which reflect the depleted core of the swirl [snapshot~4 and Fig.~\ref{fig:edge}(a)].

For the broad cone ($\beta=7\pi/6$), the confined states are polarized flocks at low and intermediate noise, circular at low noise (snapshot~6). They are several times larger than the narrow-cone states, with $\langle\delta_{\rm c.m.}\rangle_{r,t}\approx40$--$300$ at most confined points and a median of $123$, against $12$ for the narrow cone [Figs.~\ref{fig:fig6}(d)--\ref{fig:fig6}(f)]. At larger noise and large $\alpha_d$, these flocks repeatedly split into subflocks that separate and merge again (oscillating flocks, snapshot~7) or subdivide into weakly coupled, weakly polarized clusters (sub-swarms, snapshot~8). Sustained rotation is absent. Outside the confined region, groups spread superdiffusively or ballistically as elongated chains (snapshot~9), or diffusively as unstable swarms (snapshot~10). Movies of these states are provided as Supplemental Material~\cite{SM} and listed in Appendix~\ref{sec:movies}.

Swirls occur only for the narrow cone and only in variants~(ii) and~(iii), not in variant~(i) (Appendix~\ref{ssec:score_variants}). Since variants~(i) and~(iii) differ only in whether empty candidate cones compete, admitting them is what prevents rotation in variant~(i), in which an agent at the edge can turn toward an empty heading (Sec.~\ref{sec:EAM}). Swirls do not occur either when the recruitment term is removed while empty candidate cones stay excluded (Fig.~\ref{fig:c3}(g) and Appendix~\ref{sec:Scontrol}). A delay of one decision interval, in which a decision acts on the motion one step later, shifts the region of swirling-state formation toward lower noise values [Figs.~\ref{fig:c3}(a) and~\ref{fig:c3}(e)]. Furthermore, refining the decision interval toward the continuous-time limit, $\Delta t\to\Delta t/m$, $m>1$, at a fixed turning rate $\alpha_d/\Delta t$ and a fixed rotational diffusion coefficient $D_r=(\pi\eta)^2/(6\Delta t)$, half the variance $(\pi\eta)^2/3$ of the noise per unit time, maps the model at $(\alpha_d,\eta)$ onto the same model at $(\alpha_d/m,\eta/\sqrt m)$. At fixed rates, the swirls thus disappear once $\alpha_d/m$ falls below the smallest reorientation angle at which they occur, about $5\pi/12$ at $N=200$ and $\pi/4$ at $N=2000$ for three headings; for five or nine, $\pi/3$ at $N = 200$ (Appendices~\ref{sec:Nstationary} and~\ref{sec:refine}). The refined model forms swirls at $m$ times larger turning rates and diffusion coefficients. Finally, the range of reorientation angles over which swirls form depends on the action set. When the number of candidate headings is increased from three to $n_c=5$ or $9$, with the headings evenly spaced in $[-\alpha_d,\alpha_d]$, narrow-cone groups remain confined at nearly the same parameter values and still form swirls, but now throughout the entire range $\alpha_d \in [\pi/3,\,3\pi/4]$ (Appendix~\ref{sec:refine}).

\begin{figure*}
\centering
\includegraphics[width=1.0\textwidth]{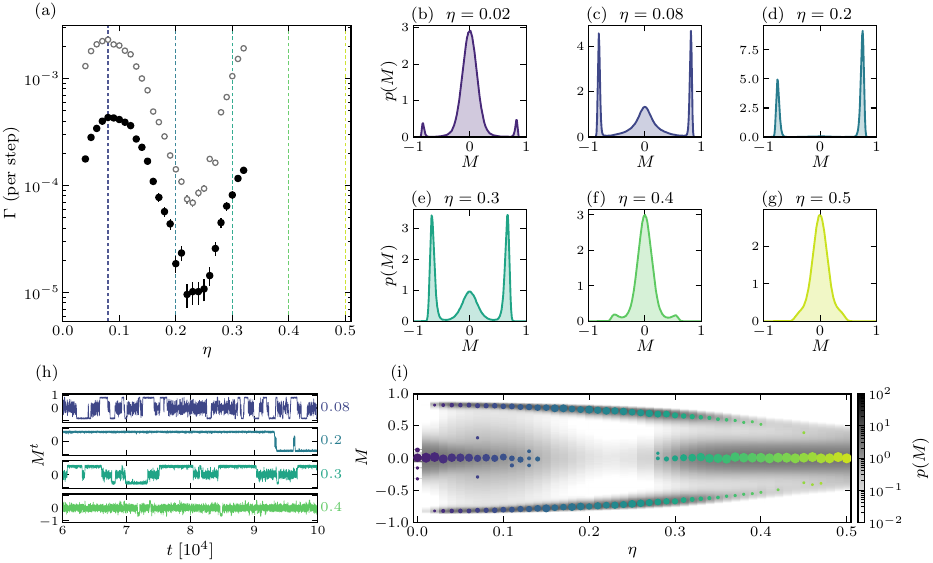}
\caption{\textbf{Reversals of the sense of rotation in the narrow-cone swirls} ($\beta=5\pi/12$, $\alpha_d=\pi/2$).
(a) Rate $\Gamma(\eta)$ of reversals between clockwise and counterclockwise rotation per time step, at the noise amplitudes of the swirls. A reversal is counted when $M^t$ passes from beyond $+M_{\rm th}$ to beyond $-M_{\rm th}$, or back, and does not return beyond the old threshold within 50 steps, for the thresholds $M_{\rm th}=M^*$ (filled circles) and $M_{\rm th}=M^*/2$ (open circles). Here $M^*$ is the position of the maximum at $M>0.1$ of the symmetrized density $[p(M)+p(-M)]/2$. $\Gamma$ is the number of reversals in all replicas divided by the number of replicas and by the length of the window, with Poisson error bars. Dashed lines mark the noise amplitudes of panels (c)--(g).
(b)--(g) Probability density $p(M)$ of the milling parameter, averaged over the replicas, at the noise amplitudes indicated.
(h) Segments of the trajectories $M^t$ of one replica, at the noise amplitudes given on the right.
(i) $p(M)$ as a function of $\eta$ (gray scale) and positions of its local maxima higher than $5\%$ of its largest one (circles), whose areas grow linearly with the probability in the corresponding basin. At $\eta=0.2$, 18 of the 25 replicas reverse at most once in the window, so that the unequal heights of the chiral peaks in (d) are a sampling effect.}
\label{fig:fig8}
\end{figure*}

These observations support the conjecture that milling in this class of models requires sufficiently large alignment errors, generated when intermediate reorientation angles are combined with either noise or a delay~\cite{Wang2023}, together with interactions that amplify them. Wide or rigid vision cones and the option of selecting an empty candidate cone dampen this amplification and thereby suppress milling, whereas finer reorientation choices promote it.

Milling in alignment-only models with a restricted field of view has also been reported for agents with a metric interaction radius in periodic domains~\cite{Costanzo2018}. This is not in conflict with our results: the periodic boundaries fix the density and may compensate the suppression of milling by rigid vision cones that we observe in open space. Related mechanisms also rely on aiming errors~\cite{Strombom2011}: with explicit attraction, blind zones promote milling~\cite{NewmanSayama2008}, noise can drive active Brownian particles from translation to rotation~\cite{Erdmann2005}, and in active colloids perception-dependent motility~\cite{Bauerle2020} and delayed attraction~\cite{Wang2023} produce rotating states. The closest analogue of Fig.~\ref{fig:fig6} is the phase diagram of a data-driven fish-school model~\cite{Calovi2014}, where attraction and alignment are separate interactions with independently tunable strengths, whereas here a single alignment rule with a restricted vision cone produces the rotation.

\subsection{Reversals of the sense of rotation}
\label{sec:swirling}

The swirls reverse their sense of rotation. Here we study these reversals using the statistics of the milling parameter $M^t$ (Fig.~\ref{fig:fig8}). 
At $\eta=0.01$, agents form a polarized flock and the probability density $p(M)$ has a single maximum at $M=0$. From $\eta=0.02$ onward, chiral maxima appear beside the central maximum at nearly full amplitude, $|M|\approx0.8$. With increasing noise, the chiral maxima grow as the central maximum weakens. At still higher noise, the chiral maxima move inward and lose weight, eventually merging with the reappearing central maximum above $\eta\approx0.4$ [Figs.~\ref{fig:fig8}(b)--\ref{fig:fig8}(g) and~\ref{fig:fig8}(i)]. We call the state a swirl where the density at the chiral maxima exceeds that at $M=0$ by at least $10\%$, i.e., for $0.04\leq\eta\leq0.32$. The data thus suggest that the transition into the swirling state is more abrupt than its disappearance at high noise. Likely, at weak noise, a rare large fluctuation is required to transform the well-aligned flock into a swirl. At strong noise, by contrast, the rotation appears to fade gradually into a less ordered, nonrotating state with a mean polarization of about $0.6$.

The rate $\Gamma$ of reversals, counted with the rule given in the caption of
Fig.~\ref{fig:fig8}, varies nonmonotonically with noise
[Fig.~\ref{fig:fig8}(a)]. Lowering the reversal threshold on $|M|$ from
$M^*$ to $M^*/2$ changes only the magnitude of $\Gamma$. Reversals are
mediated by the nonrotating state: of the 7914 reversals observed for
$0.04\leq\eta\leq0.32$, $86\%$ spend at least 100 steps with $|M^t|<0.25$
on the way from one chiral maximum to the other. The dip in $\Gamma$
coincides with the noise interval in which the probability of the
nonrotating state almost vanishes [Fig.~\ref{fig:fig8}(i)]. This is qualitatively
consistent with noise-driven hopping between three minima of the effective
potential $-\ln p(M)$, located at $M=\pm M^*$ and $M=0$. Noise-driven switching between coexisting collective states is also observed in schooling fish~\cite{Tunstrom2013} and has been analyzed with coarse-grained stochastic descriptions of the order parameter~\cite{Kolpas2007,Yates2009,Jhawar2020}.

Reversals become rare as the swirl grows. In the swirl runs of Sec.~\ref{ssec:largerN} ($\beta=5\pi/12$, $\alpha_d=\pi/2$, $\eta=0.1$) the number of reversals per replica in the statistics window falls from 20 at $N=200$ to 3.4 at $N=400$ and 0.6 at $N=600$. At $N=1000$, a single reversal occurs in the 25 replicas combined, and none at $N=2000$. In the picture of hopping between the minima of $-\ln p(M)$, the barrier between the two senses of rotation thus grows with $N$, as expected, since a reversal requires more agents to turn together.

\subsection{Dependence on group size}
\label{ssec:largerN}

\begin{figure}
\centering
\includegraphics[width=1.0\columnwidth]{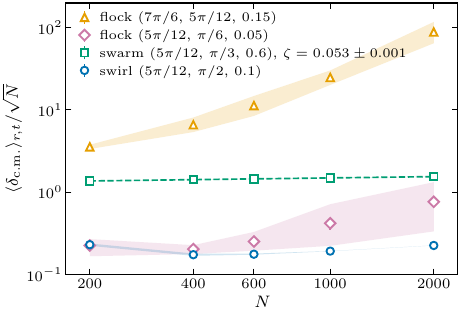}
\caption{\textbf{Growth of the group extent with the number of agents.}
Rescaled extent $\langle\delta_{\rm c.m.}\rangle_{r,t}/\sqrt N$ for the four states of Sec.~\ref{ssec:largerN}, labeled by $(\beta,\alpha_d,\eta)$. Symbols show the mean over 25 replicas (no stopping), with its standard error (mostly smaller than the symbols), and shaded bands the range between the 10th and 90th percentiles of the time averages of individual replicas. The dashed line for the swarm is a least-squares fit of $\ln[\langle\delta_{\rm c.m.}\rangle_{r,t}/\sqrt N]$ versus $\ln N$. Its slope $\zeta$, given in the legend, implies $\langle\delta_{\rm c.m.}\rangle_{r,t}\propto N^{1/2+\zeta}$. No fits are shown for the other dynamical phases because of their nonmonotonicity (swirl) or split-dominated extents (flocks). The time dependence is shown in Appendix~\ref{sec:Nstationary} and the density profiles in Appendix~\ref{ssec:boundary}. For the broad-cone flock, $t_f=4\times10^5$.}
\label{fig:growth}
\end{figure}

The effective attraction acts only on agents with an under-filled candidate cone. As shown in Sec.~\ref{ssec:pointing}, for a narrow vision cone such agents are always found at the rim of the group. As the number of agents increases, their fraction thus decreases as $N^{-1/2}$ [Appendix~\ref{ssec:boundary}], and both confinement and the overall structure of the state diagram persist [Appendix~\ref{sec:Nstationary}]. 

For large groups with broad vision cones, under-filled candidate cones may occur only sufficiently far from the group. A useful estimate is provided by the number of neighbors in the frontal vision cone of an agent located at the rim of a uniform disc and pointing away from it, $(N/\pi)(\beta-\pi+\sin\beta)$, given by Eq.~\eqref{eq:broad_rim_k}. For $\beta=7\pi/6$ and $k=7$, all candidate cones of such an agent are filled once $N\gtrsim930$. We therefore expect systems with a broad vision cone to become unstable for $N\gtrsim10^3$, splitting into several smaller groups whose sizes satisfy $(N/\pi)(\beta-\pi+\sin\beta)<k$. This expectation is confirmed in Appendix~\ref{sec:Nstationary}. The critical system size increases with $k$ and decreases with $\beta$, as documented in Figs.~\ref{fig:optimalk2} and~\ref{fig:optimalThetaRigid} of Appendix~\ref{sec:kbeta}, respectively.

To see how the individual dynamical phases evolve with increasing $N$, we followed four states for $N=200$, $400$, $600$, $1000$, and $2000$, at the same initial density for all $N$ (Sec.~\ref{sec:sim_prot}) and no stopping: three narrow-cone states, the swirl $(\beta,\alpha_d,\eta)=(5\pi/12,\pi/2,0.1)$, the swarm $(5\pi/12,\pi/3,0.6)$, and the flock $(5\pi/12,\pi/6,0.05)$, and the broad-cone flock $(7\pi/6,5\pi/12,0.15)$. 

The three narrow-cone states stay confined and keep their character at all $N$, although the polarization of the flock and the angular velocity of the swirl decrease with $N$ (Appendix~\ref{sec:Nstationary}). The latter is expected as the agents have fixed speed and the swirls are larger. The broad-cone flock is one of the few examples of confined states in the $N=2000$ broad-cone state diagram in Fig.~\ref{fig:d2k}. It grows for about $10^5$ steps at large $N$ before its extent becomes stationary, and its runs last $4\times10^5$ steps instead of $10^5$. Nevertheless, for larger $N$, it consists at almost all times of about six subgroups whose separation stays bounded (Fig.~\ref{fig:Ntime} and Table~\ref{tab:largerN}).

None of these states collapses into a dense state. For the broad-cone flock, the density decreases with increasing $N$. For the narrow vision cone, the rescaled extent $\langle\delta_{\rm c.m.}\rangle_{r,t}/\sqrt N$, which would be constant at fixed density, grows only weakly with $N$ (Fig.~\ref{fig:growth}). For the swarm, it scales as $N^\zeta$ with $\zeta=0.053\pm0.001$, increasing by $13\%$ between $N=200$ and $2000$. The swirl density rises between $N=200$ and $400$, where the swirl is still partly polarized, and then decreases by roughly $40\%$ up to $N=2000$. For the flock, the rescaled extent grows from $0.22$ at $N=200$ to $0.76$ at $N=2000$, largely due to temporary splits after which the parts merge again. When a single group contains $90\%$ of the agents, the rescaled extent at $N=2000$ is only $0.33$ (Movies~S10 and~S11 show two narrow-cone flocks of 2000 agents). A similar density selection in the absence of any metric interaction scale has been reported for metric-free swarms with projection-based interactions~\cite{PearceTurner2014}.

\section{Conclusions}
\label{sec:discussion}

In decision rules that maximize a sum of pairwise scores independent of positions, neighbor positions can favor a heading only when it modifies the perceived neighborhood, either through the recruitment term in Eq.~\eqref{eq:split}, whereby additional neighbors contribute additional score, or through the set of admissible headings, as when candidate actions without visible neighbors are excluded. The latter is a modeling choice that encodes a preference for headings with visible neighbors. The recruitment term is cohesive whenever the recruited neighbors contribute positive scores. For a finite-range vision cone, its first angular harmonic is a pointing interaction weighted by the neighbors' scores [Eq.~\eqref{eq:pointing}]. Its unconditional component, determined by the score offset, corresponds to the attraction introduced as a separate term in explicit models~\cite{Gregoire2004,Barberis2016}. With topological cones, a candidate action whose cone already contains $k$ agents cannot benefit from recruiting additional neighbors (Sec.~\ref{ssec:pointing}). Consequently, attraction acts only at the group boundary, where at least one candidate cone contains fewer than $k$ agents, and the inward bias is carried by these edge agents (Fig.~\ref{fig:edge}). A simple geometric argument, expressed by Eq.~\eqref{eq:broad_rim_k} and verified by our simulations, implies that in the scanning-alignment model the recruitment term alone confines the agents, irrespective of system size, only for narrow vision cones, whereas broad cones require sufficiently small groups or sufficiently large $k$. An action-independent neighborhood does not confine the agents. Swirls furthermore require that the dynamics amplify the turning errors induced by discrete decisions, which does not occur when empty candidate cones are preferred over misaligned ones (Sec.~\ref{sec:phase}).

Other routes from decisions to cohesion exist, such as utilities based on the path entropy of future visual states~\cite{Charlesworth2019,Turner2023}, which are not pairwise additive, and active inference~\cite{ActiveInference2022,CouzinFriston2024}. A restricted field of view is known to promote aggregation through perception-dependent motility~\cite{LavergneBechingerScience2019}, steering toward perceived neighbors~\cite{Barberis2016,Chandra2025}, self-steering combined with velocity alignment~\cite{Negi2024,Shi2025}, regulation of the visual projection of the group~\cite{Pearce2014,PearceTurner2014}, and purely vision-based interactions~\cite{BastienRomanczuk2020}.

Beyond flocking, the recruitment term of Eq.~\eqref{eq:split} applies unchanged to adaptive networks. Nodes that choose their links by maximizing an additive utility would preferentially rewire toward like-minded neighbors, providing a decision-based counterpart to the rewiring dynamics of Refs.~\cite{Holme2006,Vazquez2008}. Promising directions for future work include relating the utility structure identified here to information-theoretic measures of collective decision-making~\cite{VitelliDecisionMaking}, extending the framework to decisions with memory~\cite{holubec2026feedbackcontroldelayedinteractions}, and developing a coarse-grained description of $\argmax$ decision rules~\cite{Lama2025,VitelliDecisionMaking}.

\begin{acknowledgments}
JGB and VH were supported by Charles University (project PRIMUS/22/SCI/009). We used Claude (Anthropic) to check derivations, simulation code, and grammar in the manuscript.
\end{acknowledgments}

\section*{Data Availability Statement}
The movies, the simulation code, and the data and scripts that reproduce Figs.~\ref{fig:fig1}--\ref{fig:actions} and Tables~\ref{tab:largerN}--\ref{tab:refine}, including the controls, the runs with a finite perception range, the refinement of the decision interval, the runs with finer action sets, and the test of the update order, are available at~\cite{supplementary}. The movies are also provided as Supplemental Material~\cite{SM}. Additional data and materials are available from the authors upon reasonable request.

\appendix

\section{Movies}
\label{sec:movies}

The movies are provided as Supplemental Material~\cite{SM} and are deposited at~\cite{supplementary}. They show scanning agents with $k=7$ in the center-of-mass frame, over the last 1000 recorded frames of runs of $4\times10^4$ steps from the initial condition of Sec.~\ref{sec:sim_prot}. Movies~S1--S3 show the broad-cone flock, the swarm, and the swirl of Sec.~\ref{ssec:largerN} at $N=200$ (left) and $2000$ (right), replica~0 of the runs of Fig.~\ref{fig:growth}. Movies~S4--S9 show, at $N=200$, a chain, $(7\pi/6,\pi/12,0.2)$; a mixed flock, $(5\pi/12,\pi/3,0.075)$; an oscillating flock, $(7\pi/6,5\pi/12,0.35)$; a swarm, $(5\pi/12,\pi/12,0.35)$; a tag-like flock, $(5\pi/12,\pi/4,0.125)$; and a V-shaped flock, $(5\pi/12,\pi/2,0)$. Movies~S10 and~S11 show the narrow-cone flocks $(5\pi/12,\pi/6,0.05)$ (Sec.~\ref{ssec:largerN}) and $(5\pi/12,\pi/6,0.025)$ at $N=200$ (left) and $2000$ (right), replica~0 of the runs of Figs.~\ref{fig:fig4} and~\ref{fig:d2k}.

\section{Alternative variants of the decision rule}
\label{ssec:score_variants}

\begin{figure*}[t]
\centering
\includegraphics[width=1.0\textwidth]{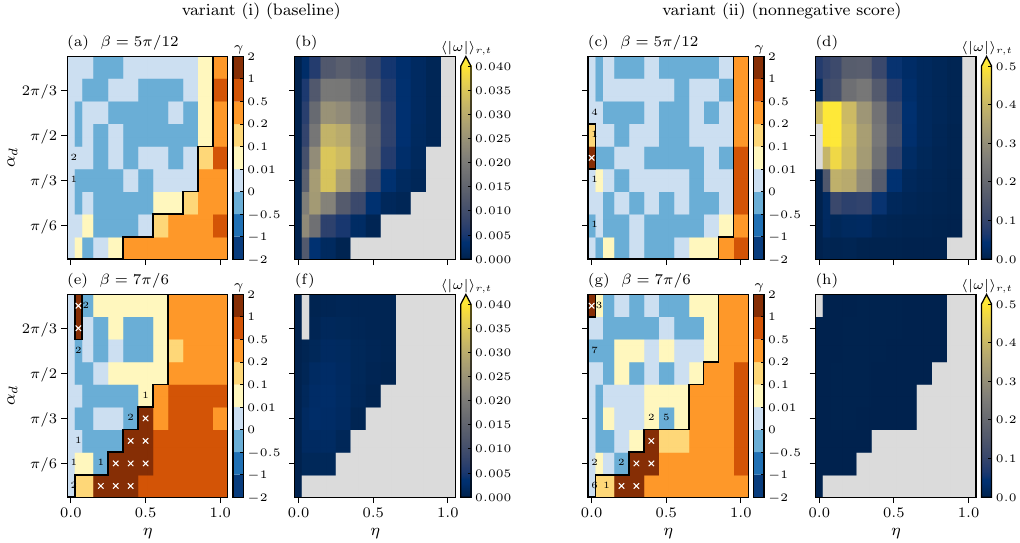}
\caption{\textbf{Variants~(i) and~(ii) of the decision rule.}
Growth exponent $\gamma$ (a,c,e,g) and absolute angular velocity $\langle|\omega|\rangle_{r,t}$ (b,d,f,h) for the baseline variant~(i) (a,b,e,f) and the nonnegative score variant~(ii) (c,d,g,h), as functions of the noise amplitude $\eta$ and the reorientation angle $\alpha_d$, for the narrow cone, $\beta=5\pi/12$ (a)--(d), and the broad cone, $\beta=7\pi/6$ (e)--(h). Crosses, numbers, and the bands of $\gamma$ are as in Figs.~\ref{fig:Sfig5} and~\ref{fig:fig4}, black lines enclose the confined points, and unconfined points are gray in the maps of $\langle|\omega|\rangle_{r,t}$. The color scale of $\langle|\omega|\rangle_{r,t}$ extends to $0.04$ for variant~(i), about an order of magnitude below the swirls of Fig.~\ref{fig:fig6}, and to $0.5$ for variant~(ii).}
\label{fig:variants}
\end{figure*}

Figure~\ref{fig:variants} repeats the stability analysis of Sec.~\ref{sec:cohesive} and the map of $\langle|\omega|\rangle_{r,t}$ of Fig.~\ref{fig:fig6} for variants~(i) and~(ii), on a coarser grid. In the baseline variant~(i), empty candidate cones have $U_i^t(\alpha)=0$ and compete with recruiting candidate cones. Its confined regions are extended for both cone angles [Figs.~\ref{fig:variants}(a) and~\ref{fig:variants}(e)], so neither the exclusion of empty candidate cones nor a nonnegative score is necessary for confinement. These regions support only flocking and swarming [Figs.~\ref{fig:variants}(b) and~\ref{fig:variants}(f)]. Large, slowly rotating groups do not occur either: by Eq.~\eqref{eq:omegadelta}, a group whose agents all move tangentially at speed $v_0$ has $\omega^t\delta^t_{\rm c.m.}=cv_0\geq v_0$, whereas here $\langle|\omega|\rangle_{r,t}\langle\delta_{\rm c.m.}\rangle_{r,t}\leq0.42v_0$.

In the nonnegative score variant~(ii), an empty candidate cone can win only through an exact tie (Sec.~\ref{sec:EAM}), so that this variant effectively excludes empty candidate cones, as variant~(iii) does by construction. It reproduces the phenomenology of variant~(iii), including swirls for the narrow cone, which rotate faster than in variant~(iii) and persist up to $\alpha_d=3\pi/4$ [Figs.~\ref{fig:variants}(c),~\ref{fig:variants}(d), and~\ref{fig:variants}(g)].

\section{Dependence on the neighbor number and the vision-cone angle}
\label{sec:kbeta}

\begin{figure*}
\centering
\includegraphics[width=1.0\textwidth]{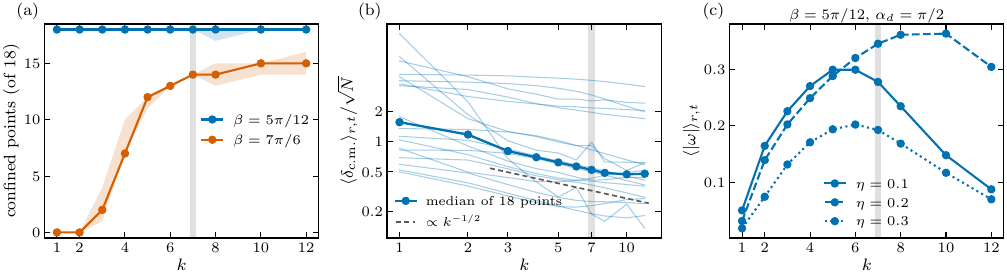}
\caption{\textbf{Dependence on the topological neighbor number $k$, from confined states only.}
(a) Number of confined points among the 18 values of $(\alpha_d,\eta)$, for the narrow and the broad cone; the bands show the range obtained with the thresholds $\gamma\leq0.05$ and $\gamma\leq0.2$ instead of $\gamma\leq0.1$. (b) Rescaled stationary extent $\langle\delta_{\rm c.m.}\rangle_{r,t}/\sqrt N$ at the 18 narrow-cone points, all of which are confined for every $k$ (thin lines), and its median (symbols); the dashed line is proportional to $k^{-1/2}$. (c) Absolute angular velocity of the narrow cone at $\alpha_d=\pi/2$. The gray band marks $k=7$.}
\label{fig:optimalk2}
\end{figure*}

Sections~\ref{sec:sim_prot}--\ref{sec:results} use $k=7$ and the cone angles $\beta=5\pi/12$ and $7\pi/6$. Figure~\ref{fig:optimalk2} shows how confinement depends on $k$ at $\alpha_d=\pi/6$, $\pi/3$, and $\pi/2$ and $\eta=0.05$, $0.1$, $0.2$, $0.3$, $0.4$, and $0.5$, classified with the criterion of Sec.~\ref{sec:cohesive}.

For the narrow cone, all 18 points are confined for every $k$; their extent decreases approximately as $k^{-1/2}$ up to $k\approx8$ and then levels off, and swirls need $k\geq3$. For the broad cone, no point is confined for $k\leq2$, and the number of confined points saturates from $k\approx6$--$7$ on, for each of the thresholds on $\gamma$. The analysis thus singles out no optimal $k$, but $k=7$ lies at the onset of the range in which confinement no longer depends on $k$, and in the range in which the narrow cone swirls; larger values would serve as well.

\begin{figure}
\centering
\includegraphics[width=\columnwidth]{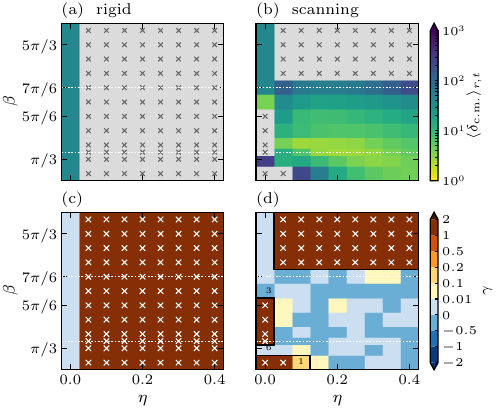}
\caption{\textbf{Dependence on the vision-cone angle.}
Time- and replica-averaged group extent $\langle\delta_{\rm c.m.}\rangle_{r,t}$ (a,b) and growth exponent $\gamma$ (c,d), for rigid (a,c) and scanning (b,d) agents, as functions of the noise amplitude $\eta$ and the vision-cone angle $\beta$, at $\alpha_d=5\pi/12$. Crosses, numbers, and the bands of $\gamma$ are as in Figs.~\ref{fig:Sfig5} and~\ref{fig:fig4}; points at which at least half of the replicas dispersed are gray in (a,b), and black lines in (d) enclose the confined points. Dotted lines mark the cone angles $\beta=5\pi/12$ and $7\pi/6$ used in the main text.}
\label{fig:optimalThetaRigid}
\end{figure}

Figure~\ref{fig:optimalThetaRigid} shows how confinement and the extent of the groups vary with $\beta$ at $\alpha_d=5\pi/12$. Rigid agents are confined only at vanishing noise, where the polarized initial state persists, for every $\beta$. Scanning agents are confined at all nonzero noise amplitudes for $\pi/3\leq\beta\leq7\pi/6$, with the most compact groups at intermediate noise for $\beta\approx5\pi/12$--$\pi/2$, and disperse at every nonzero noise amplitude for $\beta\geq4\pi/3$. The cone geometry of Sec.~\ref{ssec:pointing} accounts for the loss of confinement with $\beta$ and with decreasing $k$: at the edge of a uniform disc of 200 agents, the cone along the outward normal holds $(N/\pi)(\beta-\pi+\sin\beta)\approx1.5$ agents for $\beta=7\pi/6$ and $11.5>k$ for $\beta=4\pi/3$, so that even agents at the edge of a compact group rarely have an under-filled cone.

\begin{figure*}
\centering
\includegraphics[width=1.0\textwidth]{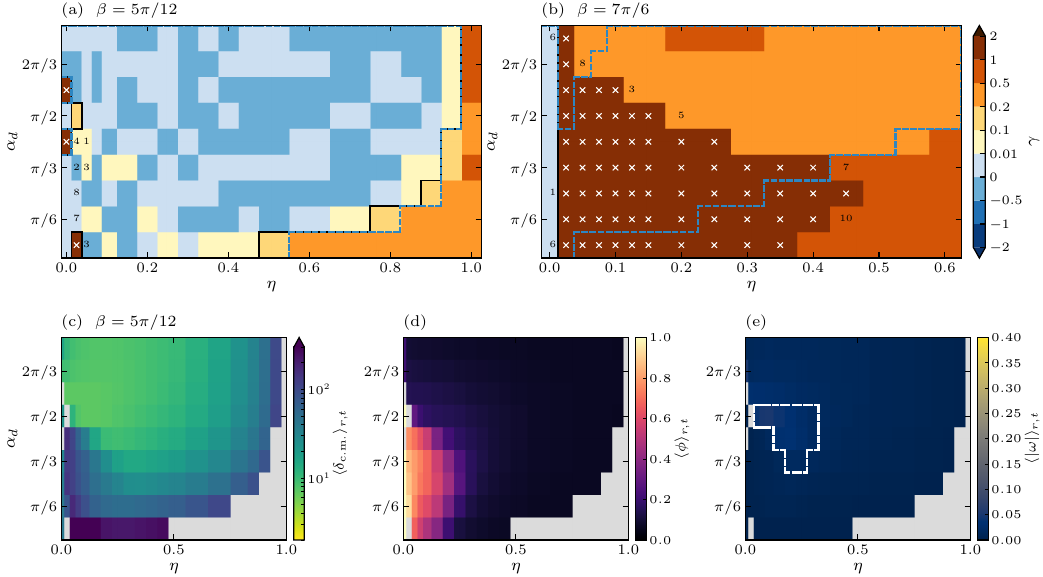}
\caption{\textbf{A single neighbor, $k=1$.} (a,b) Exponent $\gamma$, crosses, and numbers as in Fig.~\ref{fig:fig4}, for $\beta=5\pi/12$ (a) and $\beta=7\pi/6$ (b). Solid black lines enclose the points confined for $k=1$, and dashed blue lines those confined for $k=7$ [Figs.~\ref{fig:fig4}(a) and~\ref{fig:fig4}(b)]. (c)--(e) Group extent $\langle\delta_{\rm c.m.}\rangle_{r,t}$, polarization $\langle\phi\rangle_{r,t}$, and absolute angular velocity $\langle|\omega|\rangle_{r,t}$ for the narrow cone, at the points confined for $k=1$ (others gray), on the scales of Figs.~\ref{fig:fig6}(a)--\ref{fig:fig6}(c); dashed white lines in (e) enclose the points at which $k=7$ gives $\langle|\omega|\rangle_{r,t}\geq0.15$. The broad cone is confined only at $\eta=0$, where the polarized initial state persists.}
\label{fig:k1}
\end{figure*}

Figure~\ref{fig:k1} additionally repeats the stability and state diagrams of Figs.~\ref{fig:fig4} and~\ref{fig:fig6} for $k=1$. Every admissible candidate turn then recruits exactly one neighbor, the nearest agent in its cone, so that the recruitment term can favor a candidate turn only through which neighbor it recruits, not through how many. For the narrow cone, this and the exclusion of empty candidate cones suffice: the groups are confined at 152 of the 157 points confined for $k=7$, and at no other point [Fig.~\ref{fig:k1}(a)]. They are larger than for $k=7$, by a median factor of $1.5$, polarized, $\langle\phi\rangle_{r,t}\geq0.5$, only for $\alpha_d\leq5\pi/12$ and $\eta\leq0.15$, and they do not rotate, $\langle|\omega|\rangle_{r,t}<0.06$ [Figs.~\ref{fig:k1}(c)--\ref{fig:k1}(e)], consistent with the onset of swirls at $k=3$ in Fig.~\ref{fig:optimalk2}(c). For the broad cone, the groups are not confined at any nonzero noise [Fig.~\ref{fig:k1}(b)], consistent with Eq.~\eqref{eq:broad_rim_k}: at the edge of a compact group of $N=200$ agents, the cone oriented along the outward normal contains only about 1.5 agents on average.

\section{Dependence on the initial condition}
\label{sec:zeronoise}

All runs in the main text start from a fully polarized state. Repeating the stability analysis of Fig.~\ref{fig:fig4} for $\alpha_d\leq\pi/2$ with five replicas per point that start with random orientations, at random positions in the same square, reproduces the classification at nonzero noise at 112 of the 114 narrow-cone points and at 84 of the 90 broad-cone points. The differences lie at the edges of the confined regions, in particular for the broad cone at weak noise.

Without noise, the outcome depends on the initial condition throughout. From random orientations, the groups of both cones disperse in all our runs. From a fully polarized state, rigid agents keep the initial flock and its extent. Scanning agents with narrow cones either form a single, highly polarized V-shaped flock or split into several such subflocks, presumably because agents that perceive no neighbor along any candidate heading keep their heading while others follow them. For broad cones, agents almost always perceive neighbors along some candidate headings, and neither the splitting nor the V-shapes occur. At weak but nonzero noise, rare fluctuations can disrupt local order. Finite reorientation steps can then amplify this disorder, allowing groups to grow to large extents over finite times~\cite{BH2025,Turner2023,Charlesworth2019}.

\section{Recruitment channel switched off}
\label{sec:Scontrol}

\begin{figure*}
\centering
\includegraphics[width=1.0\textwidth]{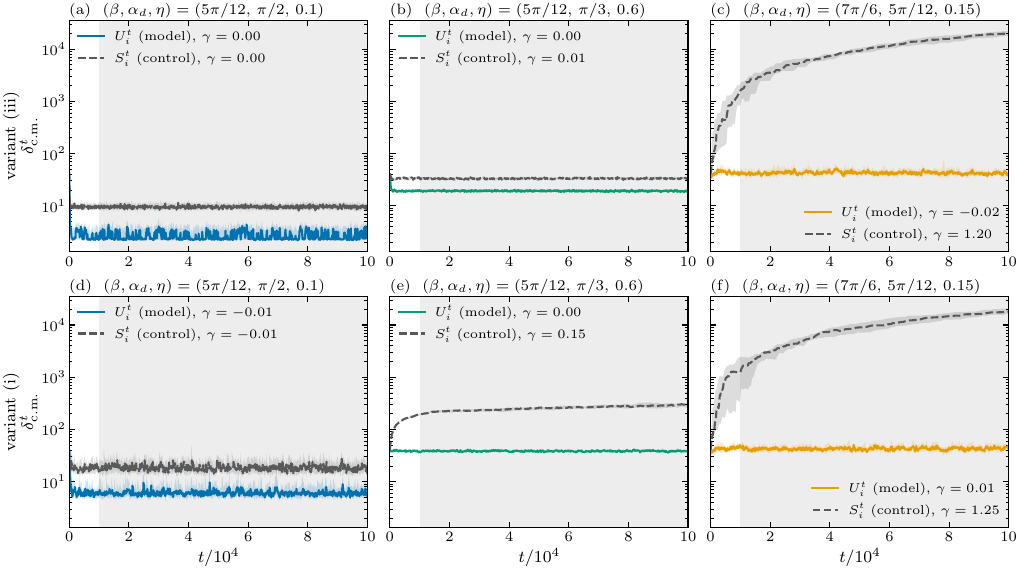}
\caption{\textbf{Control with the average score.} Group extent $\delta^t_{\rm c.m.}$ for the model, which maximizes $U_i^t$ (solid), and for the control, which maximizes $S_i^t=U_i^t/N_i^t$ (dashed), at the parameters $(\beta,\alpha_d,\eta)$ of the swirl, the swarm, and the broad-cone flock of Sec.~\ref{ssec:largerN}, for variant~(iii) (a)--(c) and variant~(i) (d)--(f). Lines show the median over 25 replicas (no stopping), and bands the interquartile range. The gray area marks the window $t\in[t_f/10,t_f]$ over which $\gamma$, given in the legends, is fitted as in Sec.~\ref{sec:cohesive}.}
\label{fig:control}
\end{figure*}

To investigate the dynamics with the recruitment term switched off, we first replace the utility $U_i^t(\alpha)$ by the average score per recruited neighbor, $S_i^t(\alpha)=U_i^t(\alpha)/N_i^t(\alpha)$, assigning $S_i^t(\alpha)=a$ to empty candidate cones (Sec.~\ref{sec:gen_model}). Since $U_i^t=kS_i^t$ wherever all candidate cones of an agent contain at least $k$ agents, the control changes decisions only in the boundary layer of Fig.~\ref{fig:edge}. In variant~(iii), empty candidate cones are excluded under either rule, so the second cohesive channel survives. Accordingly, the narrow-cone groups stay confined under the control [Figs.~\ref{fig:control}(a) and~\ref{fig:control}(b)], although they become larger, and the swirl turns into a polarized group, with $\langle\phi\rangle_{r,t}=0.65$ instead of $0.32$. The broad-cone flock, whose candidate cones are rarely empty (Fig.~\ref{fig:edge}), spreads ballistically [Fig.~\ref{fig:control}(c)].

The informative cases are variants~(i) and~(ii), which become identical under the utility $S_i^t(\alpha)$ and in which empty candidate cones remain eligible. In these variants, the normalization seemingly closes both cohesive channels and indeed abolishes the confinement of the swarm [Fig.~\ref{fig:control}(e)] and of the broad-cone flock [Fig.~\ref{fig:control}(f)]. Interestingly, at $(\alpha_d,\eta)=(\pi/2,0.1)$, where variant~(i) forms a polarized group, the group remains confined, although it becomes more extended [Fig.~\ref{fig:control}(d)]. An empty heading, with $S_i^t=0$, loses against headings along which the neighbors are, on average, aligned with the agent. In a polarized group, these correspond to the headings that keep the agent within the group. The first cohesive channel thus survives the normalization, albeit in a much weaker form. Even this residual cohesive bias is absent in rigid-cone models.

We now isolate the second cohesive channel of Sec.~\ref{sec:gen_model} by considering rigid agents (no recruitment), while a candidate heading is still admitted only if its own cone is not empty. This control confines most of the narrow-cone groups that the main model confines [Fig.~\ref{fig:c3}(c)], but they do not rotate [Fig.~\ref{fig:c3}(g)], and at the parameters of the swirl and the swarm they are larger, $\langle\delta_{\rm c.m.}\rangle_{r,t}=16$ and $28$ instead of $3.2$ and $19$. The broad-cone flock of Sec.~\ref{ssec:largerN} is not confined, and neither are the broad-cone groups at most other points [Fig.~\ref{fig:c3}(d)].

\section{Dependence on \texorpdfstring{$N$}{N} and long runs at \texorpdfstring{$N=2000$}{N=2000}}
\label{sec:Nstationary}

\begin{figure*}
\centering
\includegraphics[width=1.0\textwidth]{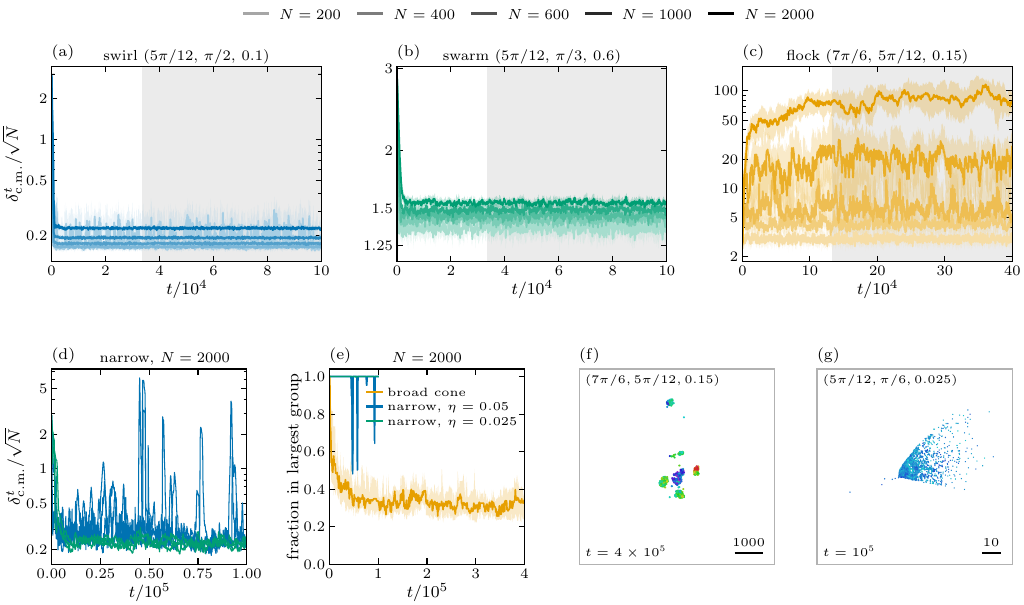}
\caption{\textbf{Time dependence of the group extent for $N=200$--$2000$, and long runs at $N=2000$.}
(a)--(c) Rescaled extent $\delta^t_{\rm c.m.}/\sqrt N$ of the swirl, the swarm, and the broad-cone flock of Sec.~\ref{ssec:largerN}, labeled by $(\beta,\alpha_d,\eta)$, for $N=200$, $400$, $600$, $1000$, and $2000$ (light to dark): median (line) and interquartile range (band) of 25 replicas (no stopping); $t_f=4\times10^5$ for the flock. The gray area marks the statistics window of Fig.~\ref{fig:growth} and Table~\ref{tab:largerN}. (d) Rescaled extent of the narrow-cone flocks $(5\pi/12,\pi/6,0.05)$ (blue) and $(5\pi/12,\pi/6,0.025)$ (green) at $N=2000$, the three replicas of Fig.~\ref{fig:d2k} at each point. (e) Fraction of the agents in the largest group, a connected component of the network that links each agent to its seven nearest agents, for the broad-cone flock of (c) at $N=2000$ (median and interquartile range) and for the narrow-cone flocks of (d). (f), (g) Configurations at the end of replica~0 of the broad-cone flock at $N=2000$ and of the narrow-cone flock at $\eta=0.025$, in the center-of-mass frame; agents are colored by their heading as in Fig.~\ref{fig:fig6}, and the scale bars give lengths in units of $v_0\Delta t$.}
\label{fig:Ntime}
\end{figure*}

\begin{figure*}
\centering
\includegraphics[width=0.9\textwidth]{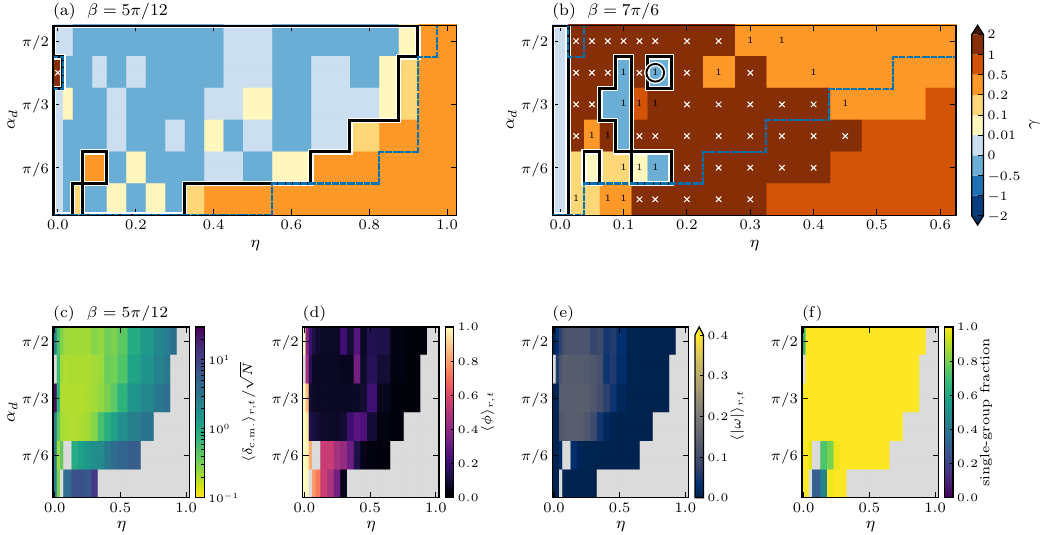}
\caption{\textbf{Stability and state diagram at $N=2000$.} (a,b) Exponent $\gamma$, crosses, and numbers as in Fig.~\ref{fig:fig4}, for $\beta=5\pi/12$ (a) and $\beta=7\pi/6$ (b). Solid black lines enclose the points confined at $N=2000$, and dashed blue lines those confined at $N=200$ (Fig.~\ref{fig:fig4}). (c)--(f) For the narrow cone, at the points confined at $N=2000$: rescaled extent $\langle\delta_{\rm c.m.}\rangle_{r,t}/\sqrt N$, polarization $\langle\phi\rangle_{r,t}$, absolute angular velocity $\langle|\omega|\rangle_{r,t}$, and the fraction of the time in which one group holds at least $90\%$ of the agents. Three replicas per point, which start in a square of side $100\sqrt{10}$. The circle in (b) marks the broad-cone flock of Sec.~\ref{ssec:largerN}.}
\label{fig:d2k}
\end{figure*}

Figure~\ref{fig:d2k} repeats the stability and state diagrams of Figs.~\ref{fig:fig4} and~\ref{fig:fig6} at $N=2000$, for $\alpha_d\leq\pi/2$, with three replicas per point. For the narrow cone, most of the points stay confined. The lost points lie at the noisiest edge of the confined region and at weak noise and small $\alpha_d$ [Fig.~\ref{fig:d2k}(a)]. Swirls occur from $\alpha_d=\pi/4$ on [Fig.~\ref{fig:d2k}(e)], and at 80 of the 86 points confined at $N=2000$, one group holds at least $90\%$ of the agents during at least $90\%$ of the statistics window [Fig.~\ref{fig:d2k}(f)]. For the broad cone, only 14 of the 61 points confined at $N=200$ stay confined, six of them at vanishing noise, where the polarized initial state persists, and the others at $\eta\leq0.15$. At 40 of the 96 points, at least two of the three replicas disperse, against 19 points at which most replicas disperse at $N=200$ [Fig.~\ref{fig:d2k}(b)].

The extents of the swirl, the swarm, and the broad-cone and narrow-cone flocks of Fig.~\ref{fig:growth} are stationary over the statistics window [Figs.~\ref{fig:Ntime}(a)--\ref{fig:Ntime}(d)], and Table~\ref{tab:largerN} lists the averages over it. Fitted as in Sec.~\ref{sec:cohesive}, over $t\in[t_f/10,t_f]$, $\gamma$ exceeds the threshold only for the broad-cone flock at $N=1000$ and $2000$, which grows for about $10^5$ steps before its extent becomes stationary [Fig.~\ref{fig:Ntime}(c)]. Nevertheless, it remains split into typically six groups of at least 50 agents, which split, wander, and merge [Figs.~\ref{fig:Ntime}(e) and~\ref{fig:Ntime}(f)]. Narrow-cone flocks of $2000$ agents form a single group most of the time [Figs.~\ref{fig:Ntime}(e) and~\ref{fig:Ntime}(g)].

\begin{table}[!t]
\caption{Averages over the statistics window and over 25 replicas (no stopping) for the four states of Sec.~\ref{ssec:largerN} ($t_f=4\times10^5$ for the broad-cone flock): polarization, absolute angular velocity, rescaled extent, the exponent $\gamma$ fitted as in Sec.~\ref{sec:cohesive}, and the fraction $f_1$ of the time in which the largest group [Fig.~\ref{fig:Ntime}(e)] holds at least $90\%$ of the agents.}
\label{tab:largerN}
\begin{ruledtabular}
\begin{tabular}{lrccccc}
state & $N$ & $\langle\phi\rangle_{r,t}$ & $\langle|\omega|\rangle_{r,t}$ & $\langle\delta_{\rm c.m.}\rangle_{r,t}/\sqrt N$ & $\gamma$ & $f_1$\\
\hline
swirl & 200 & 0.32 & 0.272 & 0.230 & $0.00$ & 1.00\\
 & 400 & 0.10 & 0.302 & 0.174 & $0.00$ & 1.00\\
 & 600 & 0.08 & 0.249 & 0.176 & $0.00$ & 1.00\\
 & 1000 & 0.08 & 0.182 & 0.192 & $0.00$ & 1.00\\
 & 2000 & 0.09 & 0.113 & 0.225 & $0.00$ & 1.00\\
\hline
swarm & 200 & 0.13 & 0.0066 & 1.37 & $0.00$ & 1.00\\
 & 400 & 0.10 & 0.0037 & 1.42 & $0.00$ & 1.00\\
 & 600 & 0.09 & 0.0025 & 1.45 & $0.00$ & 1.00\\
 & 1000 & 0.07 & 0.0016 & 1.49 & $0.00$ & 1.00\\
 & 2000 & 0.05 & 0.0008 & 1.55 & $0.00$ & 1.00\\
\hline
flock & 200 & 0.85 & 0.0025 & 3.56 & $0.00$ & 0.96\\
 & 400 & 0.82 & 0.0012 & 6.61 & $+0.01$ & 0.84\\
 & 600 & 0.79 & 0.0007 & 11.3 & $0.00$ & 0.65\\
 & 1000 & 0.71 & 0.0004 & 24.8 & $+0.14$ & 0.28\\
 & 2000 & 0.50 & 0.0001 & 88.8 & $+0.13$ & 0.002\\
\hline
narrow flock & 200 & 0.98 & 0.023 & 0.224 & $0.00$ & 1.00\\
 & 400 & 0.98 & 0.015 & 0.203 & $0.00$ & 1.00\\
 & 600 & 0.97 & 0.013 & 0.251 & $0.00$ & 1.00\\
 & 1000 & 0.96 & 0.011 & 0.418 & $0.00$ & 0.98\\
 & 2000 & 0.87 & 0.015 & 0.763 & $0.00$ & 0.90\\
\end{tabular}
\end{ruledtabular}
\end{table}

\section{Boundary layer}
\label{ssec:boundary}

\begin{figure*}
\centering
\includegraphics[width=\textwidth]{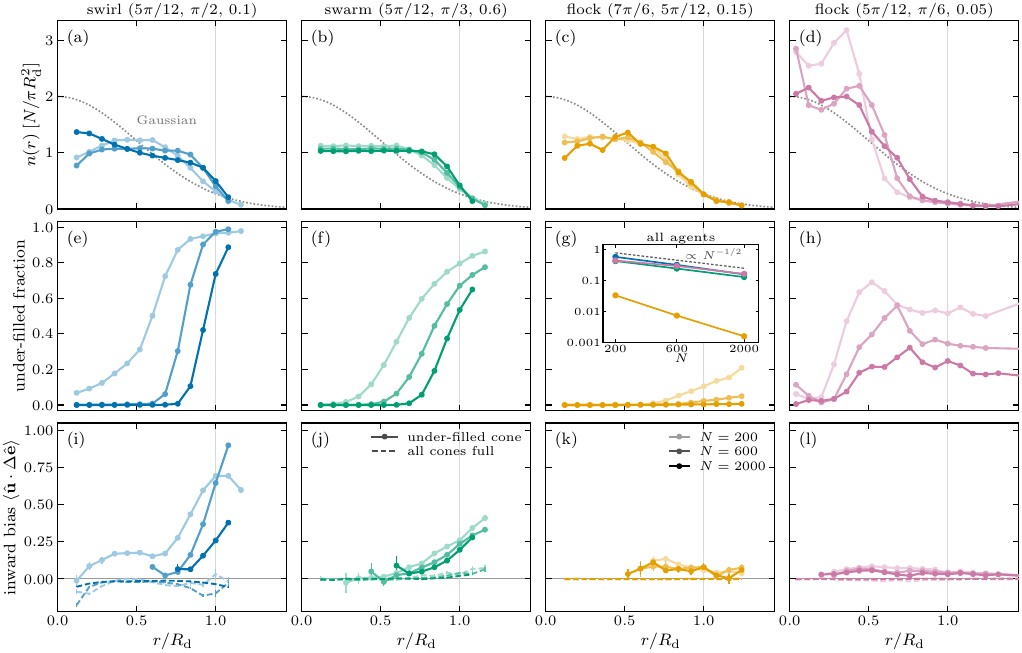}
\caption{\textbf{Boundary layer.}
Profiles versus the distance $r$ from the center of mass, in units of the radius of a uniform disc with the same extent, $R_{\rm d}=\sqrt2\,\delta^t_{\rm c.m.}$, for the swirl, the swarm, the broad-cone flock, and the narrow-cone flock of Sec.~\ref{ssec:largerN} [columns, labeled by $(\beta,\alpha_d,\eta)$] at $N=200$, $600$, and $2000$ (light to dark). At $N\geq600$, the broad-cone flock still grows at the times sampled, and at $N=2000$ it consists of several groups. (a)--(d) Density $n(r)$ in units of $N/(\pi R_{\rm d}^2)$. The dotted lines show a Gaussian profile with the same extent. (e)--(h) Fraction of agents with at least one candidate cone holding fewer than $k$ agents. Inset of (g): fraction of all agents with such an under-filled cone versus $N$ for the four states; the dashed line is proportional to $N^{-1/2}$. (i)--(l) Mean inward bias $\hat{\mathbf u}_i\cdot\Delta\hat{\mathbf e}_i$ of the scanning decision, for agents with an under-filled cone (solid) and for agents whose three cones are full (dashed). Here $\Delta\hat{\mathbf e}_i$ is the difference between the heading chosen by the scanning rule and the heading a rigid agent would choose in the same configuration, and $\hat{\mathbf u}_i$ is the unit vector from the agent to the center of mass. Averages over 25 replicas and 20 configurations per replica, $10^3$ steps apart in $t\in[1.1\times10^4,3\times10^4]$, and for the narrow-cone flock over the configurations at $t=10^4,2\times10^4,\dots,10^5$ of the runs of Fig.~\ref{fig:growth}, whose extent is stationary from $t\approx5\times10^3$ on. Bins holding fewer than $1\%$ of the agents are omitted, and in (i)--(l) also bins with fewer than 50 agents or with such agents from fewer than 10 replicas. Error bars give the standard error, with the replicas as independent samples, and thin vertical lines mark $r=R_{\rm d}$.}
\label{fig:edge}
\end{figure*}

Here, we demonstrate the localization of the effective attraction (Sec.~\ref{ssec:pointing}) on stored configurations of the four states of Sec.~\ref{ssec:largerN} (Fig.~\ref{fig:edge}). For every agent, we compute its decision under the scanning rule and the decision that a rigid agent would take in the same configuration, and compare the inward components of the two chosen headings.

The narrow-cone groups have sharp edges, near $r=R_{\rm d}$ for the swirl and the swarm and at $r\approx0.5$--$0.75R_{\rm d}$ for the flock, which has a dense core and a dilute tail, whereas the density of the broad-cone flock decreases gradually and follows the Gaussian profile near its edge [Figs.~\ref{fig:edge}(a)--\ref{fig:edge}(d)].

Agents with an under-filled cone are concentrated at the edge of the narrow-cone groups and are absent from their interior at $N\geq600$. Their fraction decreases approximately as $N^{-1/2}$ [inset of Fig.~\ref{fig:edge}(g)], as expected for a boundary layer of fixed width in units of the distance between agents [Figs.~\ref{fig:edge}(e)--\ref{fig:edge}(h)]. These agents carry the inward bias, whereas agents whose cones are all full have none [Figs.~\ref{fig:edge}(i)--\ref{fig:edge}(l)]. That individuals at the boundary of a group behave differently from those in its bulk has also been observed in starling flocks~\cite{Attanasi2015,Cavagna2013}.

In the broad-cone flock, under-filled cones are rare at all distances from the center and become rarer as $N$ grows [Fig.~\ref{fig:edge}(g)], as the cone geometry of Sec.~\ref{ssec:pointing} predicts. Averaged over all agents, the inward bias is therefore negligible.

\section{Finite perception range}
\label{sec:range}

The vision cones of the model carry no metric cutoff: an agent perceives its $k$ nearest neighbors within a cone regardless of their distance. To test how strongly confinement relies on this property, we introduce a perception range $R$, so that only agents within distance $R$ can enter a cone and the agent interacts with the $k$ nearest among them. We consider the parameter sets corresponding to the swirl, the swarm, and the broad-cone flock of Sec.~\ref{ssec:largerN}, with $N=200$. The groups first form with unbounded cones, during $2\times10^4$ steps from the initial condition of Sec.~\ref{sec:sim_prot}, and then evolve for $10^5$ steps with the finite range. Since an agent with no other agent within $R$ keeps its heading up to the noise, the extent $\delta^t_{\rm c.m.}$ then mostly records such stragglers, and Table~\ref{tab:range} gives instead the size $N_{\rm cl}$ of the largest proximity cluster. We measure $R$ in units of $\ell$, the median distance of the $k$th neighbor in the cone along the heading in the group with unbounded cones.

\begin{table}[!t]
\caption{Finite perception range $R$, at the parameters of the swirl, the swarm, and the broad-cone flock of Sec.~\ref{ssec:largerN} with $N=200$: number $N_{\rm cl}$ of agents in the largest proximity cluster after $10^5$ steps with the finite range, median and, in parentheses, range over eight replicas; its rate of change $\dot N_{\rm cl}$ over the second half of these steps, in agents per $10^4$ steps, mean and standard error over the replicas (dashes where the group had dissolved before); and the median number $N_0$ of agents with no other agent within $R$. The length $\ell$ is $1.26$, $9.44$, and $16.1$ for the swirl, the swarm, and the flock.}
\label{tab:range}
\begin{ruledtabular}
\begin{tabular}{lrrccr}
state & $R$ & $R/\ell$ & $N_{\rm cl}$ & $\dot N_{\rm cl}$ & $N_0$\\
\hline
swirl & 2 & 1.6 & 1 (1--1) & -- & 200\\
 & 4 & 3.2 & 1 (1--1) & -- & 200\\
 & 8 & 6.3 & 1 (1--2) & -- & 200\\
 & 16 & 12.7 & 2 (1--123) & $-21\pm5$ & 195\\
 & 32 & 25.4 & 191 (185--196) & $-1.0\pm0.2$ & 9\\
 & 64 & 50.8 & 199 (196--200) & $-0.09\pm0.05$ & 1\\
\hline
swarm & 15 & 1.6 & 190.5 (185--196) & $-0.8\pm0.2$ & 7.5\\
 & 30 & 3.2 & 200 (200--200) & $0$ & 0\\
 & 60 & 6.4 & 200 (200--200) & $0$ & 0\\
\hline
flock & 25 & 1.6 & 55.5 (41--70) & $-1.2\pm0.6$ & 15\\
 & 50 & 3.1 & 74 (51--113) & $-1.6\pm2.5$ & 0\\
 & 100 & 6.2 & 104.5 (64--124) & $-6.2\pm2.2$ & 0\\
 & 200 & 12.4 & 158.5 (100--200) & $+0.9\pm1.6$ & 0\\
 & 400 & 24.8 & 172 (108--200) & $-6.4\pm3.2$ & 0\\
\end{tabular}
\end{ruledtabular}
\end{table}

The swarm keeps all its agents for $R\geq3\ell$ and loses them only slowly at $R=1.6\ell$. The swirl dissolves for $R\leq13\ell$, agent by agent, although the range then exceeds the diameter of the swirl, $2R_{\rm d}\approx9$, and keeps its agents only at $R=51\ell$. The broad-cone flock splits into polarized subgroups that move apart, in most replicas at all ranges tested (Table~\ref{tab:range}).

The difference lies in how far an agent, or a subgroup, travels before its heading decorrelates. With uniform angular noise of amplitude $\eta$, the heading of an agent that perceives no one decorrelates over $-1/\ln[\sin(\pi\eta)/(\pi\eta)]$ steps, about $6/(\pi\eta)^2$ for small $\eta$. In the swarm, at $\eta=0.6$, this is 1.5 steps, so that a straying agent diffuses and rarely leaves the range. In the swirl, at $\eta=0.1$, it is 61 steps, so that a straying agent moves ballistically over about 60 step lengths, comparable to the range $R=64$ ($51\ell$) at which the swirl keeps its agents. The polarized subgroups of the broad-cone flock presumably keep their common heading even longer. Confinement thus requires a range larger than the distance over which agents or subgroups move persistently.

\section{Refinement of the decision interval and of the action set}
\label{sec:refine}

\begin{figure*}[t]
\centering
\includegraphics[width=1.0\textwidth]{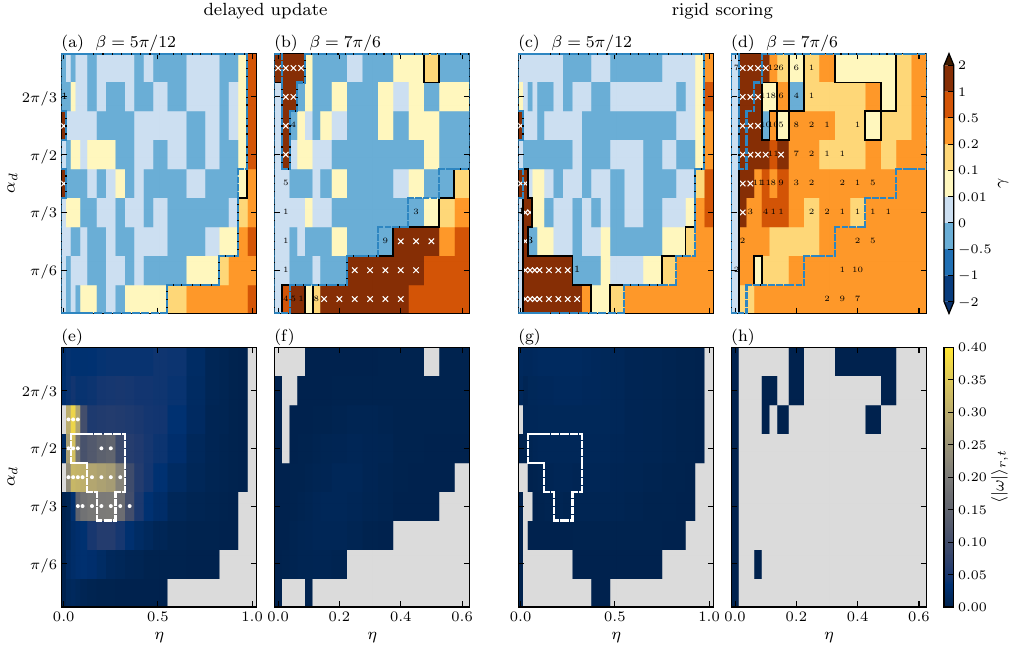}
\caption{\textbf{State diagrams with a delayed update and with rigid scoring.} (a)--(d) Exponent $\gamma$, crosses, and numbers as in Fig.~\ref{fig:Sfig5}, and (e)--(h) absolute angular velocity $\langle|\omega|\rangle_{r,t}$ at the confined points (others gray), on the scale of Fig.~\ref{fig:fig6}(c), for the delayed update of Appendix~\ref{sec:refine} (a,b,e,f) and the rigid-scoring control of Appendix~\ref{sec:Scontrol} (c,d,g,h), with $\beta=5\pi/12$ (a,c,e,g) and $\beta=7\pi/6$ (b,d,f,h). The runs have the initial conditions and random numbers of the runs of Figs.~\ref{fig:fig4} and~\ref{fig:fig6} at the same parameters. Solid black lines enclose the confined points, and dashed blue lines the points confined with the standard rule [Figs.~\ref{fig:fig4}(a) and~\ref{fig:fig4}(b)]. White dots mark swirls at $\eta>0$, by the criterion of Sec.~\ref{sec:swirling} applied to $M^t$ recorded every 100 steps, and dashed white lines enclose the points at which the standard rule gives $\langle|\omega|\rangle_{r,t}\geq0.15$ [Fig.~\ref{fig:fig6}(c)].}
\label{fig:c3}
\end{figure*}

We test the mapping of Sec.~\ref{sec:phase}, $(\alpha_d,\eta)\to(\alpha_d/m,\eta/\sqrt m)$, for the swirls of Table~\ref{tab:refine}. Each run with $m>1$ starts from the final configuration of the run at $m=1$, a swirl, with all positions multiplied by $m$, and lasts as long in physical time: $2\times10^4$, $1.5\times10^4$, and $1.2\times10^4$ original steps for $N=200$, $1000$, and $2000$.

\begin{table}[!t]
\caption{Refinement of the decision interval by a factor $m$ at fixed turning rate and rotational diffusion, for the swirls at $(\beta,\alpha_d,\eta)=(5\pi/12,\pi/2,\eta_0)$, so that the refined model has $\alpha_d=\pi/(2m)$ and $\eta=\eta_0/\sqrt m$. The runs at $m=1$ start from the polarized initial condition, those at $m>1$ from the swirl formed at $m=1$. Polarization, milling parameter, and extent and absolute angular velocity in units of the original step and time step, averaged over the last two thirds of the runs and over the replicas, and the number of replicas that rotate, $\langle|M|\rangle_t>0.5$.}
\label{tab:refine}
\begin{ruledtabular}
\begin{tabular}{rrrccccc}
$N$ & $\eta_0$ & $m$ & $\langle\phi\rangle_{r,t}$ & $\langle|M|\rangle_{r,t}$ & $\langle\delta_{\rm c.m.}\rangle_{r,t}$ & $\langle|\omega|\rangle_{r,t}$ & swirls\\
\hline
200 & 0.1 & 1 & 0.29 & 0.56 & 3.1 & 0.30 & 4/4\\
 &  & 1.25 & 0.61 & 0.34 & 6.1 & 0.19 & 1/4\\
 &  & 1.5 & 0.87 & 0.12 & 4.1 & 0.06 & 0/4\\
 &  & 2 & 0.94 & 0.07 & 3.0 & 0.05 & 0/4\\
 &  & 3 & 0.97 & 0.10 & 3.1 & 0.07 & 0/4\\
 &  & 4 & 0.98 & 0.07 & 1.0 & 0.07 & 0/4\\
\hline
200 & 0.2 & 1 & 0.11 & 0.74 & 2.7 & 0.36 & 4/4\\
 &  & 1.25 & 0.12 & 0.79 & 2.1 & 0.47 & 4/4\\
 &  & 1.5 & 0.59 & 0.36 & 5.5 & 0.19 & 1/4\\
 &  & 2 & 0.80 & 0.19 & 9.6 & 0.04 & 0/4\\
 &  & 3 & 0.80 & 0.23 & 12.4 & 0.04 & 0/4\\
 &  & 4 & 0.84 & 0.24 & 13.6 & 0.03 & 0/4\\
\hline
1000 & 0.1 & 1 & 0.07 & 0.82 & 6.0 & 0.18 & 2/2\\
 &  & 2 & 0.09 & 0.93 & 3.2 & 0.37 & 2/2\\
 &  & 3 & 0.91 & 0.13 & 19.6 & 0.03 & 0/2\\
\hline
2000 & 0.1 & 1 & 0.09 & 0.81 & 10.1 & 0.11 & 2/2\\
 &  & 2 & 0.12 & 0.90 & 4.7 & 0.25 & 2/2\\
 &  & 3 & 0.54 & 0.52 & 9.4 & 0.16 & 1/2\\

\end{tabular}
\end{ruledtabular}
\end{table}

Swirls of 200 agents do not survive the refinement beyond $m=1.25$--$1.5$, also when the runs start from the initial condition of Sec.~\ref{sec:sim_prot}. For larger $m$, the agents form polarized groups, which stay confined. The value $\langle|M|\rangle_{r,t}\approx0.2$ of those at $\eta_0=0.2$ comes from their asymmetric shape (Sec.~\ref{sec:sim_prot}) and from turns of the whole group without a net sense of rotation. Swirls of 1000 and 2000 agents survive $m=2$ but mostly not $m=3$. A swirl that survives keeps its extent and its angular velocity in units of the new step: at $m=2$, its extent in units of the original step halves, and its angular velocity per original step doubles (Table~\ref{tab:refine}). The size of the swirls is thus set by the step length rather than by the turning rate and the noise, and at fixed $N$, a swirl that survived the limit of continuous decisions would shrink to a point. Larger groups tolerate a finer decision interval, but nothing in the data indicates that this scaling breaks down for large groups. The swarm at $(5\pi/12,\pi/3,0.6)$, in contrast, survives the refinement: its images for $m=2$ and $4$ are confined and disordered, with extents of $21$ and $38$ original step lengths instead of $19$.

The swirls also depend on the order of the update. If the positions are advanced with the heading before the decision, so that a decision acts on the motion one step later, which amounts to a reaction delay of one decision interval, the swirls move to weaker noise, down to $\eta=0.025$, and extend to $\alpha_d=7\pi/12$ [Figs.~\ref{fig:c3}(a) and~\ref{fig:c3}(e)], whereas the confined regions hardly change [Figs.~\ref{fig:c3}(a) and~\ref{fig:c3}(b)]. Equation~\eqref{eq:Pos_upd} advances the position with the new velocity, as in Ref.~\cite{ChateH-2008}, where this choice is expected not to matter for the asymptotic behavior of Vicsek-type models. Here, it matters for the swirls. The swarm at $(5\pi/12,\pi/3,0.6)$ stays confined and disordered with either order, but its extent changes from $19.3$ to $14.0$.

\begin{figure*}[t]
\centering
\includegraphics[width=1.0\textwidth]{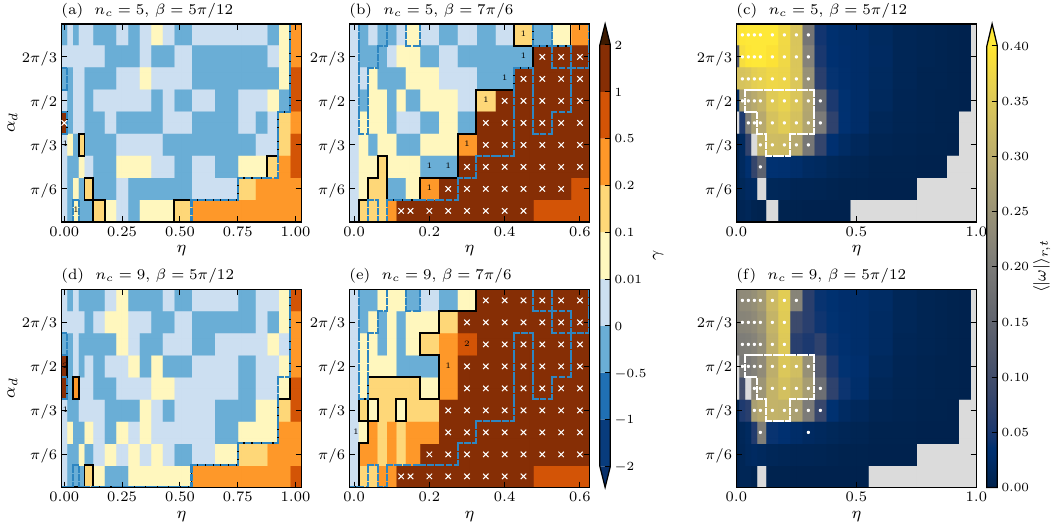}
\caption{\textbf{Finer action sets.} Stability and swirls with $n_c=5$ (a)--(c) and $n_c=9$ (d)--(f) candidate headings evenly spaced in $[-\alpha_d,\alpha_d]$, with five replicas per point. (a,b,d,e) Exponent $\gamma$, crosses, and numbers as in Fig.~\ref{fig:Sfig5}, for $\beta=5\pi/12$ (a,d) and $\beta=7\pi/6$ (b,e); solid black lines enclose the confined points, and dashed blue lines the points confined with three headings in the same five replicas. (c,f) $\langle|\omega|\rangle_{r,t}$ for $\beta=5\pi/12$ at the confined points (others gray), on the scale of Fig.~\ref{fig:fig6}(c); white dots mark swirls at $\eta>0$ by the criterion of Sec.~\ref{sec:swirling}, and dashed white lines enclose the points at which three headings give $\langle|\omega|\rangle_{r,t}\geq0.15$.}
\label{fig:actions}
\end{figure*}

Finally, the agents of the model choose among three candidate headings. To test this restriction, we repeated the stability and state diagrams of Figs.~\ref{fig:fig4} and~\ref{fig:fig6} with $n_c=5$ and $9$ candidate headings evenly spaced in $[-\alpha_d,\alpha_d]$, so that the largest turn is unchanged and smaller turns become available, with ties going to the smallest turn. Each point has five replicas, with the initial conditions and random numbers of the first five replicas of Figs.~\ref{fig:fig4} and~\ref{fig:fig6} (Fig.~\ref{fig:actions}). For the narrow cone, the confined region hardly changes: it holds 152 and 155 points for $n_c=5$ and $9$, against 155 for three headings with the same replicas [Figs.~\ref{fig:actions}(a) and~\ref{fig:actions}(d)]. The swirls persist, but they now form at low and intermediate noise for all $\alpha_d$ from $\pi/3$ to $3\pi/4$, and at single points for $\alpha_d=\pi/4$, with $\langle|\omega|\rangle_{r,t}$ up to $0.43$ and $0.36$ [Figs.~\ref{fig:actions}(c) and~\ref{fig:actions}(f)]. For $\alpha_d\geq7\pi/12$, where three headings give no swirls, the agents in the swirls choose the largest turn in fewer than $5\%$ of the decisions and otherwise mostly keep their heading or take the smallest turn. The window of turns of $75^\circ$ to $90^\circ$ in Sec.~\ref{sec:phase} is thus a property of the rule with three headings. For the broad cone, finer action sets make the groups more polarized and confine them at fewer points, 82 and 43 against 96, mainly at larger noise and, for $n_c=9$, also at $\alpha_d\leq5\pi/12$ [Figs.~\ref{fig:actions}(b) and~\ref{fig:actions}(e)].

\bibliography{references}

\end{document}